\documentclass[aps,prd,preprint,a4paper,showpacs,nofootinbib,superscriptaddress]{revtex4-2}
\usepackage{bm}
\usepackage{indentfirst}
\usepackage{amsmath}
\usepackage{graphicx}
\usepackage{amsmath}
\usepackage{CJK}

\usepackage{amssymb}
\usepackage{subfigure}
\usepackage{amssymb}
\usepackage{epstopdf}
\usepackage[section]{placeins}

\usepackage[utf8]{inputenc}

\usepackage{color}
\usepackage[T1]{fontenc}
\usepackage{txfonts}
\usepackage{orcidlink}

\usepackage{booktabs}
\usepackage{array}
\usepackage{appendix}
\usepackage{hyperref}
\hypersetup{
    colorlinks=true,
    linkcolor=red,
    citecolor=blue,
}

\begin{document}

\title{Non-linearly scalarized supermassive black holes}

\author{Shoupan Liu}
\email{shoupan_liu@163.com}
\address{\textit{Center for Gravitation and Cosmology, College of Physical Science and Technology, Yangzhou University, Yangzhou 225009, China}}

\author{Yunqi Liu}
\email{yunqiliu@yzu.edu.cn (corresponding author)}
\address{\textit{Center for Gravitation and Cosmology, College of Physical Science and Technology, Yangzhou University, Yangzhou 225009, China}}

\author{Yan Peng}
\email{yanpengphy@163.com}
\address{\textit{School of Mathematical Sciences, Qufu Normal University, Qufu, Shandong 273165, China}}

\author{Cheng-Yong Zhang}
\email{zhangcy@email.jnu.edu.cn}
\address{\textit{Department of Physics and Siyuan Laboratory, Jinan University, Guangzhou 510632, China}}

\baselineskip=0.5 cm

\begin{abstract}
In this study, we investigate a nonlinear mechanism driving the formation of scalarized rotating black holes within a scalar-Gauss-Bonnet gravity framework that includes an additional squared Gauss-Bonnet term.
With the specific coupling function, Kerr metric is a solution to this modified gravity.
In linear level Kerr black holes are stable against the scalar perturbation, while nonlinearly they suffer the so-called ``nonlinear scalarization" and  are unstable.
By employing a pseudo-spectral method, we derive the spectrum of nonlinearly scalarized rotating black hole solutions, revealing multiple scalarized branches. 
Our analysis demonstrates that both the black hole's spin and the additional squared Gauss-Bonnet term significantly influence the existence and properties of these solutions. 
Furthermore, we explore the thermodynamic properties of nonlinearly scalarized rotating black holes, and find that the scalarized black holes are entropically favored over Kerr black holes of the same mass and spin across a wide range of parameters.
\end{abstract}

\maketitle

\newpage

\section{Introduction}
\label{Introduction}

The study of black holes has been a cornerstone of gravitational physics since the formulation of general relativity (GR). 
In GR, black holes are elegantly simple, defined solely by their mass, angular momentum, and electric charge \cite{kerr2009kerr}, as encapsulated by no-hair theorems \cite{PhysRevLett.34.905,PhysRevLett.26.331,doi:10.1142/S0218271815420146} and the Kerr hypothesis\cite{Herdeiro2023}, implying that black holes possess no additional degrees of freedom. 
However, unresolved questions in gravitational physics, such as the nature of dark matter, dark energy and the behavior of gravity in extreme regimes, have motivated the exploration of modified gravities. 
In the recent decades, significant modified gravity frameworks admit the existence of additional degree of freedom defying the Kerr hypothesis, so that the stationary vacuum spacetimes need not to be described by the Kerr metric.
Popular examples encompass black holes interacting with conformally coupled scalar fields \cite{Bekenstein:1974sf}, Skyrme matter \cite{Luckock:1986tr,Droz:1991cx}, Yang-Mills fields \cite{Volkov:1989fi, Bizon:1990sr,Greene:1992fw,Maeda:1993ap}, and dilatonic black holes in Einstein-dilaton-Gauss-Bonnet theory \cite{Torii:1996yi,Kanti:1996gs,Kleihaus:2015aje,Kleihaus:2011tg,Guo:2008hf}. 

Recently, many studies have focused on extended scalar-tensor theories \cite{Antoniou:2021zoy,Liu:2022eri,Jiang:2023yyn,Zhang:2022cmu,Doneva:2022yqu,Doneva:2022ewd,Doneva:2023kkz,Liu:2022fxy,Zhang:2023jei,ZouPhysRevD.108.084007,Jiang:2023yyn,Minamitsuji:2023uyb}, which allows for the development of scalarization.
Their attractive feature is that they are perturbatively equivalent to GR in weak-field regimes, whereas in strong-field environments, second-order phase transitions can trigger the emergence of scalarized compact object states that is known as {\it spontaneous scalarization} \cite{PhysRevLett.120.131103, Silva:2017uqg,PhysRevLett.120.131102,Herdeiro:2020wei, Eichhorn:2023iab, PhysRevLett.125.231101, PhysRevD.102.104027, PhysRevLett.126.011104}. The underlying mechanism typically emerges in models featuring real scalar fields with non-minimal couplings to source terms.
These couplings manifest as effect mass-squared terms in the scalar field's dynamical equations, where negative values may trigger tachyonic instability.
The coupling sources generally include various geometric or matter invariants \cite{Herdeiro:2018wub,Zhang:2021edm,Brihaye:2018bgc,Herdeiro:2019yjy,Doneva:2017bvd, Silva:2017uqg, Antoniou:2017acq, Cunha:2019dwb, Dima:2020yac, Herdeiro:2020wei, Berti:2020kgk, Corelli:2022pio, Corelli:2022phw}.
The scalarization may follow a linear or nonlinear route, if one choice different coupling function. It is interesting to point out that the nonlinear behavior typically occurs in the theories with higher-order coupling functions \cite{Doneva:2021tvn,Blazquez-Salcedo:2022omw,Zhang:2021nnn,Liu:2022fxy,Zhang:2024spn,Gonzalez:2024ifp,Jiang:2023yyn}, where both the coupling function and its first derivative with respect to the scalar field vanish at a stationary point where its second derivative does not.
A significant number of studies focus on the nonlinear instability \cite{Doneva:2021tvn, ZouPhysRevD.108.084007, Blazquez-Salcedo:2022omw}.
As a result, this nonlinear instability leads to the formation of new black holes with scalar hair, and the spectrum of solutions is more complicated and more than one scalarized branch can exist.

In conventional framework, scalarization occurs for Kerr black holes below a critical mass scale, and also neutron stars \cite{Antoniou:2021zoy, Liu:2022eri,Jiang:2023yyn,Zhang:2022cmu,Doneva:2022yqu,Doneva:2022ewd}.
This means that the instability is unattainable for supermassive black holes. 
Reference \cite{Eichhorn:2023iab} presents the first example of a theory
in which both the Gauss-Bonnet term and an additional squared Gauss-Bonnet term are quadratically coupled to a scalar field. 
By exploring the model in spherical symmetry, they found the instability, and non-uniqueness are limited to a 
finite range of black hole masses that can be chosen to be supermassive. 
Those findings are particularly interesting, as they suggest the potential observation window through future gravitational wave detectors like LISA, Taiji, and TianQin to the underlying gravitational theory \cite{Danzmann:1997hm,Hu:2017mde,TianQin:2015yph,Li:2024rnk}.
In Ref. \cite{Liu:2025mfn}, the authors extended these investigations to rotating black holes, by employing a quadratic-exponential coupling function, they examined how black hole spin and coupling constants influence the existence and properties of these solutions. 
And Ref.\cite{Thaalba:2025ljh} explored whether this model can emerge from an effective field theory.

The primary purpose of this work is to derive the nonlinearly scalarized rotating solutions within the novel theoretical framework introduced in Ref. \cite{Eichhorn:2023iab}.
Since in the frame of nonlinear instability, the spectrum of solutions is more complicated, and exist more than one scalarized branch, it is interesting to investigate how black hole spin and the additional squared Gauss-Bonnet term influence the parameter space where scalarization occurs, whether the rotating supermassive BH still exists in large parameter space, we also will investigate the thermodynamic properties of these scalarized configurations compared to Kerr black holes.
The structure of the paper is as follows. Sec.\ref{Theoretical setup} describe the theoretical framework, the ansatz and boundary conditions. We also analyze the nonlinear scalar perturbation on the Kerr black hole in the theory. Sec.\ref{Numerical method and quantities of interest} specifies the Chebyshev pseudo-spectral numerical method which we utilize in this work and introduce physical quantities in compacted radial coordinates. Sec.\ref{Numerical Results} are devoted to exploring the parameter space to characterize the nonlinearly scalarized black hole solutions and their dependence on spin and the additional squared Gauss-Bonnet term. Finally, we summarize our results in Sec.\ref{Conclusions}. Throughout this work, we adopt units $G$ = $c$ = 1.

\section{Theoretical setup}
\label{Theoretical setup}

\subsection{Nonlinear instability of Kerr black hole}

We examine a class of four-dimensional scalar-Gauss-Bonnet gravity theory that incorporates an additional squared Gauss-Bonnet term. This theory is characterized by the action referenced in Ref. \cite{Eichhorn:2023iab} as
\begin{eqnarray}\label{Act}
    S=\frac{1}{16\pi} \int d^4x \sqrt{-g} [R-(\partial \phi)^2 + \alpha_1 F(\phi) \mathcal{G} - 2\alpha^3_2 F(\phi)(\psi \mathcal{G} - \frac{\psi ^2}{2})],
\end{eqnarray}
where $F(\phi)$ is the coupling function and $\mathcal{G}$ is the Gauss-Bonnet invariant defined as
\begin{eqnarray}\label{GB term}
   \mathcal{G}=R^2-4R_{\alpha \beta}R^{\alpha \beta}+R_{\alpha \beta \gamma \sigma}R^{\alpha \beta \gamma \sigma}.
\end{eqnarray}
The theoretical framework incorporates two coupling constants $\alpha_1$ and $\alpha_2$ which carry dimensions of length squared, a dimensionless real scalar field $\phi$, and an auxiliary field $\psi$, which has the same dimensions as $\mathcal{G}$.  

Varying the action (\ref{Act}) with respect to the metric $g_{\mu \nu}$, we derive the equations of motion written in Einstein-like form as
\begin{eqnarray}\label{EOM}
    \mathcal{E_{\mu \nu}} = G_{\mu \nu} - T_{\mu \nu} = 0,
\end{eqnarray}
where the contributions of scalar fields and curvature couplings encapsulated in the effective energy-momentum tensor $T_{\mu \nu}$:
\begin{eqnarray}\label{EM}
    T_{\mu \nu} = \nabla_{\mu} \phi \nabla_{\nu} \phi - \frac{1}{2}g_{\mu \nu}\left[(\nabla \phi)^2 + \alpha^3_2 \psi^2 F(\phi) \right] + 4 P_{\mu \alpha \nu \beta} 
    \nabla^{\alpha} \nabla^{\beta} \left[(\alpha_1 - 2\alpha^3_2 \psi) F(\phi)\right].
\end{eqnarray}
In this expression, the tensor $P_{\mu \nu \alpha \beta}$ is the double dual Riemann tensor defined as
\begin{eqnarray}\label{P}
   P_{\mu \nu \alpha \beta} \equiv \frac{1}{4} \epsilon_{\mu \nu \gamma \delta} R^{\rho \sigma \gamma \delta} \epsilon_{\rho \sigma \alpha \beta} 
   = R_{\mu \nu \alpha \beta} + 2g_{\mu [\beta}R_{\alpha]\nu} + 2g_{\nu[\alpha}R_{\beta]\mu} + R g_{\mu[\alpha}g_{\beta]\nu}.
\end{eqnarray}

The scalar field equations, obtained by varying the action (\ref{Act}) with respect to real scalar field $\phi$, takes the form
\begin{eqnarray}\label{KG}
    \square \phi = - \left [\alpha_1 \mathcal{G} - 2 \alpha^3_2 \left (\psi \mathcal{G} - \frac{\psi^2}{2} \right )\right ] \frac{F'(\phi)}{2},
\end{eqnarray}
where $F'(\phi)$ denotes the derivative of $F(\phi)$ with respect to $\phi$. 

The auxiliary field $\psi$ is acts as a Lagrange multiplier
and satisfies the constraint
\begin{eqnarray}\label{LM}
    \psi - \mathcal{G} = 0.
\end{eqnarray}
This constraint reveals that the action (\ref{Act}) effectively includes a term proportional to $\mathcal{G}^{2}$.

The theory described by the action (\ref{Act}) supports the solution of GR with the vanishing scalar $\phi=0$ and coupling function $F(\phi)=0$. To further explore the property of stability, one can substitute the equation of motion Eq.(\ref{LM}) into Eq.(\ref{KG}), and then linearize it around $\phi=0$. 
We obtain the equation of motion governing the linear scalar perturbation $\phi_p$ as follows
\begin{eqnarray}\label{perturbation Eq}
   \square \phi_p =m^2_{eff} \phi_p, ~~~~~m^2_{eff}=-\frac{1}{2} \left( \alpha_1 \mathcal{G}-\alpha_2^3 \mathcal{G}^2 \right)  F''(0) 
\end{eqnarray}
where $m^2_{eff}$ is the effective mass squared of the perturbation.
In this work, we consider the coupling function to take the form $F(\phi)= \frac{1}{\kappa}(1-e^{-\kappa \phi^{4}})$, where $\kappa$ is a coupling parameter. 
With the properties $F(0) = 0$, $F'(0) = 0$, and $F''(0) = 0$,
the equation of the scalar perturbation Eq.(\ref{perturbation Eq}) reduces to $\square \phi_p = 0$, which means at the linear level the effective mass $m^2_{eff} = 0$ and the system do not suffer instability.
However, recent nonlinear analysis by \cite{PhysRevD.106.104027} shows an intriguing result: sufficiently large initial perturbation can induce nonlinear instabilities in Kerr spacetime and it leads to the formation of nonlinearly scalarized rotating black hole.

\subsection{Equations of motion and boundary behaviors}
We are interested in stationary and axially symmetric spacetimes. In general, such spacetimes are described by a Lewis-Papapetrou-type ansatz \cite{wald1984general}, which satisfies the circularity condition and involves four unknown metric functions. In this work, we adopt a version of this ansatz as introduced in \cite{fernandes2023new}, with the line element
\begin{eqnarray}\label{LineElement}
    ds^2=-f \mathcal{N}^2 dt^2 + \frac{g}{f} \left [h \left (dr^2 + r^2 d \theta ^2 \right ) + r^2 sin^2 \theta \left (d \varphi - \frac{W}{r}(1 - \mathcal{N})dt
     \right )^2 \right ],
\end{eqnarray}
where $\mathcal{N}  = 1 - r_H / r$ regularizes the event horizon at $r=r_H$, and the dimensionless functions $f$, $g$, $h$, and $W$ depend on the radial $r$ and angular parameters $\theta$. The scalar fields $\phi$ and $\psi$ are also the function of coordinates $r$ and $\theta$ only. The three ``quasi-isotropic'' spherical coordinates $r,~\theta,~\phi$ range over the intervals
\begin{eqnarray}\label{three spatial intervals}
    r \in [r_H, \infty), \quad \theta \in [0, \pi], \quad \varphi \in [0, 2\pi].
\end{eqnarray}

Assuming the metric induced by line element (\ref{LineElement}) is a solution of the Einstein field equations Eq.(\ref{EOM}), we can derive a system of nonlinear partial differential equations (PDEs) governing the four unknown metric functions $f$, $g$, $h$, $W$ and two scalar fields $\phi$, $\psi$. In order to solve the system, we employ the following equations that diagonalize the Einstein tensor parts with respect to $\hat{\mathcal{O}} f$, $\hat{\mathcal{O}} g$, $\hat{\mathcal{O}} h$ and $\hat{\mathcal{O}} W$ (the operator $\hat{\mathcal{O}} = \partial^2_r + r^{-2} \partial^2_{\theta}$), respectively,
\begin{subequations}\label{CEinsteinFieldEq}
\begin{align}
   ~-\mathcal{E}^{\mu}_{\ \mu} + 2 \mathcal{E}^{t}_{\ t} + \frac{2W r_H}{r^2} \mathcal{E}^{\varphi}_{\ t}  &= 0,\label{CEinsteinFieldEq-1}\\
   ~\mathcal{E}^{\varphi}_{\ t} &= 0, \label{CEinsteinFieldEq-2}\\ 
   ~\mathcal{E}^{r}_{\ r} + \mathcal{E}^{\theta}_{\ \theta} &= 0,\label{CEinsteinFieldEq-3}\\
   ~\mathcal{E}^{\varphi}_{\ \varphi} - \frac{W r_H}{r^2} \mathcal{E}^{\ \varphi}_{t} - \mathcal{E}^{r}_{\ r} - \mathcal{E}^{\theta}_{\ \theta} &= 0.\label{CEinsteinFieldEq-4}
\end{align}
\end{subequations}
Together with Eq.(\ref{KG}) and Eq.(\ref{LM}), we will solve those equations in the following sections.

Moreover, the equations of motion should be subject to appropriate boundary conditions.
Asymptotic flatness imposes boundary conditions at spatial infinity ($r \rightarrow \infty$):
\begin{eqnarray}\label{infinity BC of r}
   f = g = h = 1, ~~~ 2 r_H r^2 \partial_{r} W + \chi \left( 2 r_H + r^2 \partial_{r} f \right)^2 = 0, ~~~ \text{and} ~~~ \phi = \psi = 0.
\end{eqnarray}
The boundary condition of unknown function $W$ induced from the behavior of the metric components at $r \rightarrow \infty$, where
\begin{equation}\label{metric components at inf}
 \begin{aligned}
   &g_{tt} = -f \mathcal{N}^2+\frac{g(1-\mathcal{N})^2 W^2}{f} \sin^2{\theta} = -1 + \frac{2M}{r} + \cdots, \\
   &g_{t\varphi} = -\frac{g r (1-\mathcal{N}) W}{f} \sin^2{\theta} = -\frac{2J}{r} \sin^2{\theta} + \cdots. 
 \end{aligned}
\end{equation}
One can read off the Arnowitt-Deser-Misner (ADM) mass $M$ and the angular momentum $J$ from it as
\begin{eqnarray}\label{ADM mass and angular momentum}
   M = r_H + \frac{r^2}{2}\partial_{r}{f}\Big|_{r\rightarrow\infty},  ~~~~~~~~~   J = -\frac{r_H r^2}{2} \partial_{r}{W}\Big|_{r\rightarrow\infty}.
\end{eqnarray}
Defining the dimensionless spin parameter $ \chi \equiv J/M^2$, one can readily get the boundary condition of unknown function $W$.

At the event horizon $r=r_H$, we derived the following boundary conditions by substituting the series expansions of the unknown functions into the equations of motion:
\begin{eqnarray}\label{horizon BC of r}
    f-r_H \partial_{r}f  = g + r_H \partial_{r}g  = \partial_{r}h  = W + r_H \partial_{r}W/2=0, ~~~ \text{and} ~~~ \partial_{r}{\phi}=\partial_{r}\psi=0.
\end{eqnarray}

Axial symmetry and regularity require vanishing angular derivatives at the symmetry axis:
\begin{eqnarray}\label{axial BC}
\partial_{\theta}f=\partial_{\theta}g=\partial_{\theta}h=\partial_{\theta}W=\partial_{\theta}\phi=\partial_{\theta}\psi=0 ~~~ \text{for} ~~~ \theta=0,~\pi.
 \end{eqnarray}
 
Furthermore, the absence of conical singularities imposes that on the symmetry axis
\begin{eqnarray}\label{conical singularity} 
   h = 1 ~~~ \text{for} ~~~ \theta=0,~\pi.
\end{eqnarray}

\section{Numerical Method and Quantities of Interest}
\label{Numerical method and quantities of interest}

In this section, we provide a brief overview of the numerical methods employed. We utilize the Chebyshev pseudo-spectral method in combination with the Newton-Raphson method to solve the coupled system defined by Eqs.\eqref{KG}, \eqref{LM} and \eqref{CEinsteinFieldEq}. This approach has been extensively applied in the study of rotating black hole solutions, as demonstrated in works such as \cite{fernandes2023new} and \cite{ZouPhysRevD.108.084007}.

First, we introduce the compactified radial coordinate $x=1-2r_H/r$, which maps the semi-infinite domain $r\in\left [r_H, \infty \right )$ to the finite interval $x\in\left [-1, 1 \right ]$. In this compactified coordinate system, the boundary conditions at horizon ($x=-1$) are expressed as follows
\begin{align}\label{BChorizon}
f - 2 \partial_x f =0,&\quad g + 2 \partial_x g =0,\quad \partial_x h =0,\notag\\
\quad W -  \partial_x W =0, &\quad \partial_{x} \phi =\partial_x \psi =0.
\end{align}
At spatial infinity ($x=1$) the asymptotic boundary conditions become
\begin{align}\label{BCinfinity}
     f = g = h = 1,\quad  \partial_x W + \chi (1 + \partial_x f)^2=0 \quad \text{and} \quad \phi = \psi = 0.
\end{align}

By exploiting equatorial reflection symmetry, under which $\theta \to \pi -\theta$, we limit the angular domain to $\theta \in \left[0,\pi /2 \right]$ and develop spectral expansions for the six functions $\mathcal{F}^{(k)} = \left\{f, g, h,W,\phi,\psi\right\}$. These expansions utilize tensor products of Chebyshev polynomials and cosine basis functions, expressed as follows
\begin{eqnarray}\label{ExpandSerise}
    \mathcal{F}^{(k)}\left(x,\theta\right) = \sum_{i = 0}^{N_x -1} \sum_{j = 0}^{N_{\theta} -1} \alpha^{(k)}_{ij} T_{i}(x) \cos(2j\theta),
\end{eqnarray}
$T_i(x) = \cos(i \arccos x)$ represents the $i$-th Chebyshev polynomial, $\alpha^{(k)}_{ij}$ are the spectral coefficients, and $N_x$ and $N_{\theta}$ denote the resolutions in the radial and angular directions, respectively. For the computations in this study, we primarily use resolutions of $N_x$ = 40 and $N_{\theta}$ = 8, with details on convergence provided in Appendix \ref{Appendix A}.

The system of nonlinear PDEs given by Eqs. (\ref{CEinsteinFieldEq}), Eq. (\ref{KG}) and Eq. (\ref{LM}) involves the unknown functions $\mathcal{F}^{(k)}$ along with their first and second derivatives, specifically ($\mathcal{F}^{(k)},\partial_x \mathcal{F}^{(k)},\partial^2_x \mathcal{F}^{(k)},\partial_{\theta} \mathcal{F}^{(k)}$, 
\\ 
$\partial^2_{\theta} \mathcal{F}^{(k)}, \partial_{x \theta} \mathcal{F}^{(k)}$). By expressing the PDEs in the residual form $\mathcal{R}(x,\theta,\partial \mathcal{F}^{(k)})=0$ and substituting the spectral expansions Eq.~\eqref{ExpandSerise} into these residuals, we evaluate the resulting equations at the Chebyshev-Gauss nodes, which are defined as:
\begin{eqnarray}
    x_l = \cos\left[\frac{(2l+1)\pi}{2N}\right], \quad \theta_m = \frac{(2m+1)\pi}{4N}, \quad l,m = 0,...,N-1.
\end{eqnarray}
Incorporating with the boundary conditions, also evaluated at those nodes, this yields a system of $N_{\mathcal{F}} \times N_{x} \times N_{\theta}$ algebraic equations for the spectral coefficients $\alpha^{(k)}_{ij}$, expressed as:
\begin{eqnarray}\label{Coeffs}
    \alpha^{(k)}_{ij} = \frac{4}{N_x N_{\theta}} \sum_{l = 0}^{N_x -1} \sum_{m = 0}^{N_{\theta} -1} \mathcal{F}^{(k)}(x_l,\theta_m) T_i(x_l) \cos(2j\theta_{m}).
\end{eqnarray}

To solve the algebraic equations using the Newton-Raphson iterative method, an initial guess is required. By specifying black hole parameters $r_H,~\chi,~M$, we expand the functions of the Kerr solution in the spectral series given by Eq.\eqref{ExpandSerise} to obtain the spectral coefficients $\alpha^{(k)}_{ij, Kerr}$. These coefficients $\alpha^{(k)}_{ij, Kerr}$ serve as initial guess for the Newton-Raphson solver, which is used to search the spectral coefficients $\alpha^{(k)}_{ij}$ of the prospective scalarized black hole solution. Moreover, the Newton-Raphson iteration is terminated when the residual is reduced below $10^{-10}$, ensuring the solution satisfies the field equations to within numerical precision.

Once the nonlinearly scalarized rotating black hole solution is obtained, one can extract its physically meaningful quantities. Most of them are determined by the metric functions and scalar field, evaluated either at the horizon or at spatial infinity. In the previous section, we showed how to derive the ADM mass $M$ and angular momentum $J$ from the asymptotic behavior of the metric functions, as presented in Eq. (\ref{ADM mass and angular momentum}). In the compactified coordinate system, the ADM mass take the form
\begin{eqnarray}
    M=r_H\left( 1+\partial_{x}f\right)\Big|_{x=1}.
\end{eqnarray}
Additionally, the solutions possess a scalar charge $Q_s$, which is determined by the leading term in the far-field asymptotics of the scalar field $\phi$:
\begin{eqnarray}
   \phi = \frac{Q_s}{r}+ \mathcal{O}  (r^{-2}).
\end{eqnarray}

We are also interested in the thermodynamic properties of nonlinearly scalarized rotating black holes. The stationary and axial symmetry metric (\ref{LineElement}) admits two Killing vector fields, $k = \partial_{t}$ and $\Phi = \partial_{\varphi}$, and in particular the linear combination
\begin{eqnarray}
    \xi = \partial_{t} + \Omega_H \partial_{\varphi},
\end{eqnarray}
where the horizon angular velocity $\Omega_H$ is fixed by the value of metric function at the horizon $\Omega_H = W |_{x=-1} / r_H$.
The surface gravity $\zeta$ is defined as $\zeta^2 = -1/2(\nabla_{\mu} \xi_{\nu})(\nabla^{\mu} \xi^{\nu})$, so that the Hawking temperature follows as
\begin{eqnarray}\label{Hawking T}
    T_{H} = \frac{\zeta}{2 \pi} = \frac{1}{2 \pi r_H} \frac{f}{\sqrt{gh}} \Bigg|_{x=-1}.
\end{eqnarray}
The induced metric on the horizon, 
\begin{eqnarray}
    d \Sigma^2 = \gamma_{ij} dx^i dx^j = \frac{r_H^2 g}{f} \left[ h d\theta^2 +\sin^2 {\theta}d\varphi^2 \right] \Big|_{x=-1},
\end{eqnarray}
encodes the two-dimensional geometry used to compute the entropy via the Iyer-Wald formalism \cite{wald1993black,iyer1994some}:
\begin{align}\label{entropy}
   S = \frac{A_H}{4} + \frac{1}{4} {\int_{\mathcal{H}}^{}  \,d^{2}x} \ \sqrt[]{\gamma} \left (\alpha_1 - 2\alpha^3_2 \psi \right )F(\phi) \widetilde{R},
\end{align}
where $\mathcal{H}$ denotes the two-surface, $\gamma$ is the determinant of the induced metric and $\widetilde{R}$ is the induced Ricci scalar. 
The horizon area $A_H$ is explicitly given by
\begin{align}\label{TA}
    & A_H = 2 \pi r^{2}_H\int_{0}^{\pi} \,d\theta \sin \theta \frac{g \enspace \sqrt[ ]{h}}{f}\Bigg | _{x=-1}.
\end{align}
All of these quantities are evaluated at the compactified horizon location $x=-1$.

\section{Numerical results}
\label{Numerical Results}
In this section, we present numerical results for nonlinearly scalarized rotating black hole with the coupling constants $\alpha_1=1$ throughout. We aim to investigate nonlinearly scalarized rotating solutions in the theory \eqref{Act}, especially how the dimensionless spin parameters $\chi$ and coupling constant $\alpha_2$ affect the black hole scalarization.
To achieve this, we analyze the domain of existence, focusing on three specific values of $\kappa = 6$, $\kappa = 25$ and $\kappa = 100$. These values correspond to three distinct topological solutions identified in \cite{Doneva:2021tvn}.

\subsection{The results for \texorpdfstring{$\kappa$}{kappa} = 6}

\begin{figure}[htbp]
    \centering
    \begin{minipage}{0.49\textwidth}
        \centering
        \includegraphics[width=\textwidth]{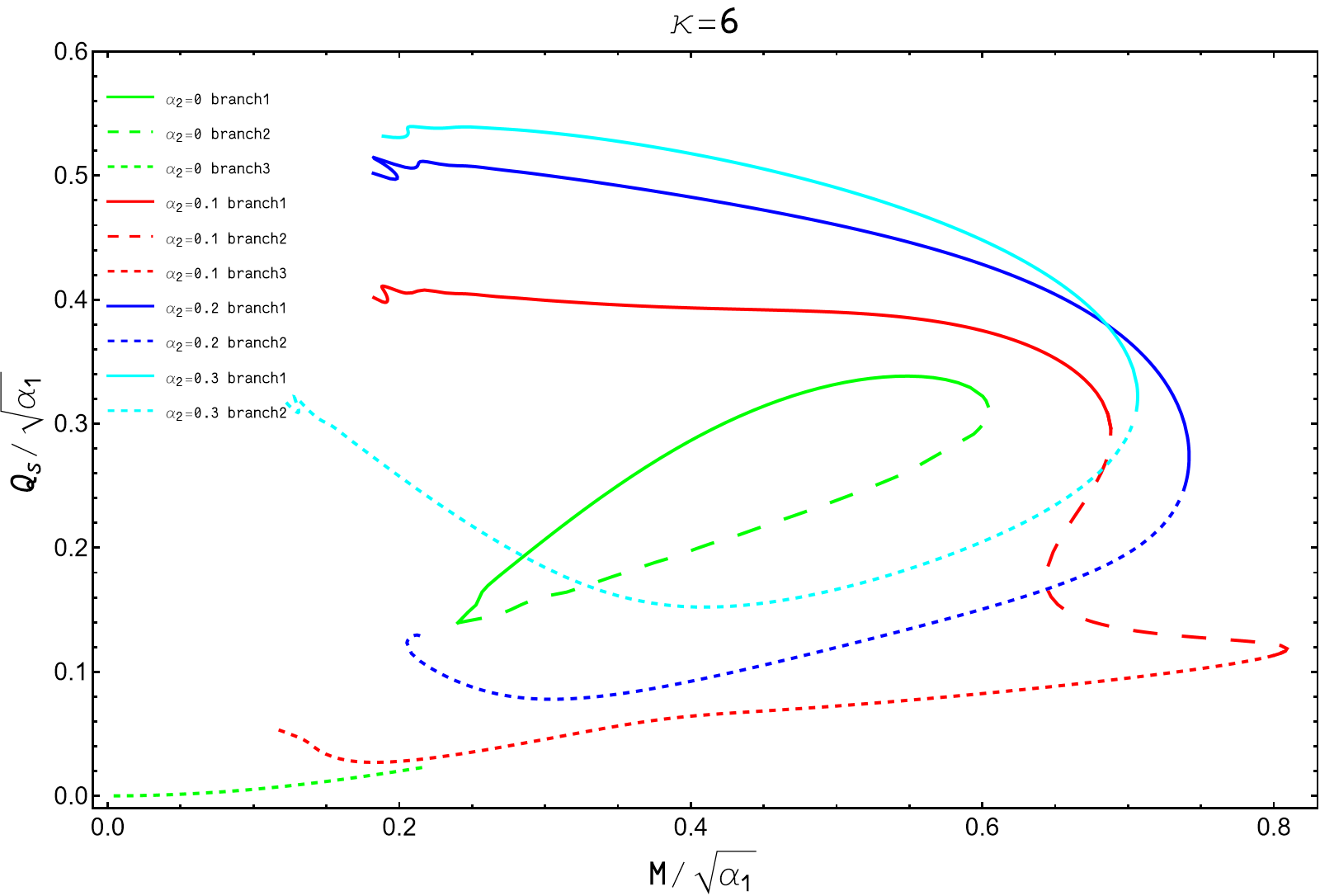}
    \end{minipage}
    \hspace{0.1cm}
    \begin{minipage}{0.49\textwidth}
	   \centering
	   \includegraphics[width=\textwidth]{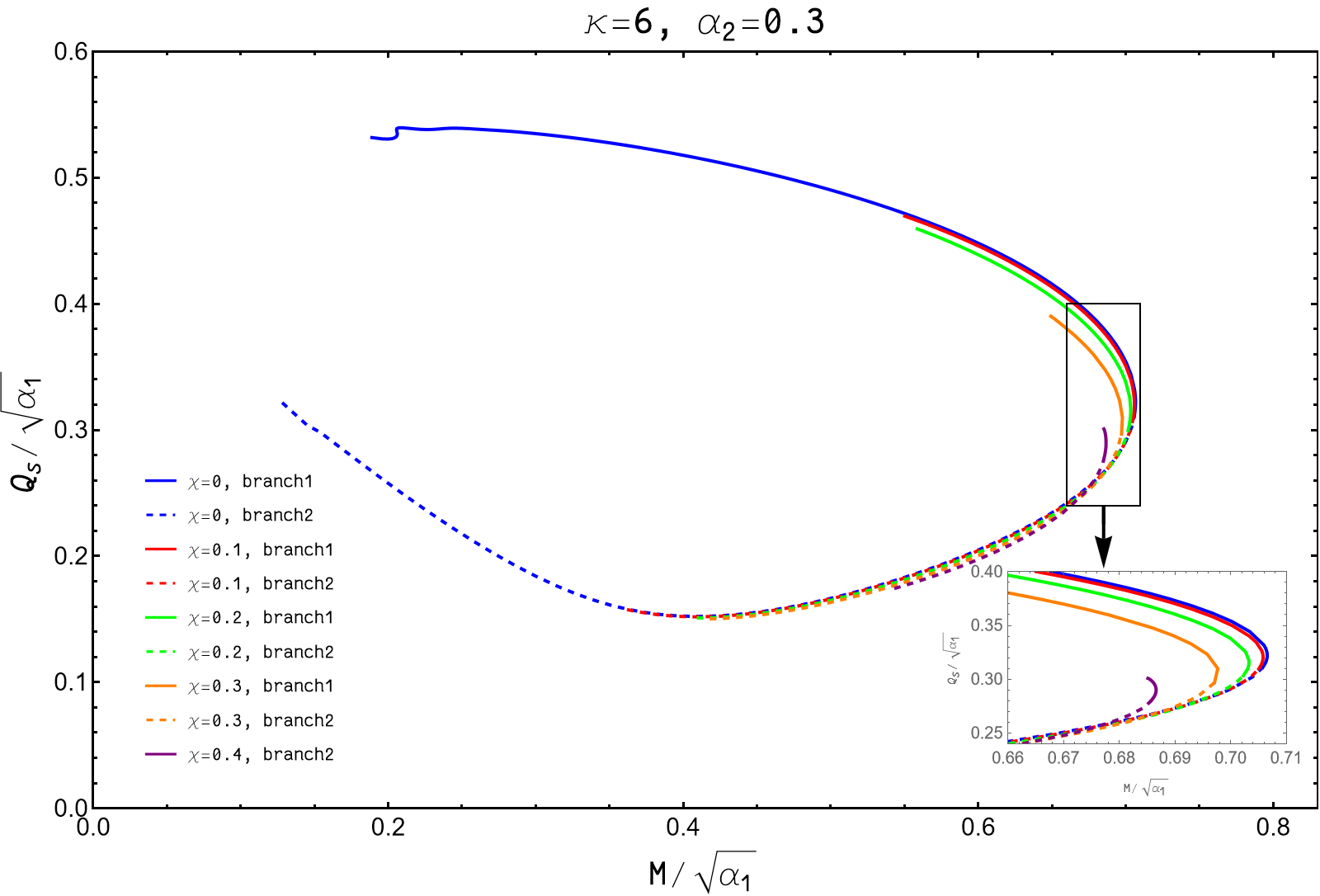}
    \end{minipage}
    \caption{The scalar charge, $Q_s/\sqrt{\alpha_1}$, plotted against the scaled ADM mass, $M/\sqrt{\alpha_1}$, for non-rotating solutions with $\kappa = 6$, $\alpha_2 = 0,~0.1,~0.2,~0.3$ (left panel), and for rotating solutions with $\alpha_2 =0.3$ (right panel).}
    \label{fig:kappa_6 Qs}
\end{figure}

We begin our analysis with the case of $\kappa = 6$. To provide a reference for the rotating case and to compare our result with those in \cite{Doneva:2021tvn}, we first consider the non-rotating ($\chi = 0$) solutions. 
The left panel in Fig. \ref{fig:kappa_6 Qs} shows the scalar charge $Q_s/\sqrt{\alpha_1}$ of spherically symmetric solutions as a function of the scaled ADM mass $M/\sqrt{\alpha_1}$ for $\alpha_2$ equal to 0, 0.1, 0.2 and 0.3. Different colors correspond to different values of $\alpha_2$, while line styles distinguish among the various branches. As shown, three distinct branches are present for $\alpha_2=0$ and $\alpha_2=0.1$, while only two branches appear for $\alpha_2=0.2$ and $\alpha_2=0.3$.

The influence of the additional $\mathcal{G}^2$ term on the domain of existence is quantified in Table \ref{tab:DeltaM_kappa_kappa6_chi=0}, which lists the mass range $\Delta M = M_{max} - M_{min}$ for the non-rotating case, where $M_{max}$ and $M_{min}$ denote the maximum and minimum of the black holes masses for a given spin, respectively. 
For this coupling strength ($\kappa=6$), the data in Table \ref{tab:DeltaM_kappa_kappa6_chi=0} reveal a clear non-monotonic dependence of $\Delta M$ on the coupling $\alpha_2$. 
The domain is most extended at $\alpha_2 = 0.1$ ($\Delta M/\sqrt{\alpha_1} \approx 0.69187$), significantly broader than in the absence of the $\mathcal{G}^2$ term ($\alpha_2 = 0$,  $\Delta M/\sqrt{\alpha_1} \approx 0.60395$).
A further increase of $\alpha_2$ to 0.2 leads to a notable contraction of the domain ($\Delta M/\sqrt{\alpha_1} \approx 0.55970$), which then expands slightly again at $\alpha_2 = 0.3$ ($\Delta M/\sqrt{\alpha_1} \approx 0.58655$). This complex behavior demonstrates that for $\kappa=6$, the additional squared term does not have a universally suppressive effect, but can non-trivially reshape space in a way that is highly sensitive to the specific coupling strength.
\setlength{\tabcolsep}{4mm}  
\begin{table}[htb]
    \centering
    \caption{The mass range $\Delta M/\sqrt{\alpha_1}$ for non-rotating ($\chi=0$) scalarized black hole with $\kappa = 6$.}
    \label{tab:DeltaM_kappa_kappa6_chi=0}
    \begin{tabular}{lcccccc}  
        \specialrule{1pt}{0pt}{0pt}  
        $\alpha_2$          & 0 & 0.1 & 0.2 & 0.3   \\
        \specialrule{0.5pt}{0em}{0em}  
        $\Delta M$   & 0.60395 & 0.69187 & 0.55970 & 0.58655   \\
        \specialrule{1pt}{0pt}{0pt}  
    \end{tabular}
\end{table}

We next consider the effect of rotation. A clear trend is observed: the presence of the $\mathcal{G}^2$ term enables scalarization of black holes with higher spin. 
For $\alpha_2=0$, solutions could only be obtained up to a maximum of $\chi=0.1$. 
In contrast, for $\alpha_2=0.1$, $0.2$ and $0.3$, rotating solutions were found up to $\chi \geq 0.4$. 
To illustrate the impact of spin, the right panel of Fig. \ref{fig:kappa_6 Qs} shows the results for a fixed $\alpha_2 = 0.3$. 
This value is selected because it provides a clear representation of the $\mathcal{G}^2$ term's influence while retaining substantial domain of existence. 
Different colors and line styles distinguish spin values $\chi$ and branches.
As $\chi$ increases, the domain of existence gradually shrinks. 
At $\chi = 0.4$, the upper branch is severely diminished, persisting only over a very narrow mass range. 
The quantitative data illustrating this shrinkage of the domain for $\alpha_2=0.1$ and $\alpha_2=0.2$ are summarized in Table \ref{tab: kappa=0.6_alpha_2=all} in Appendix \ref{Appendix B}. 
This suggests that the $\mathcal{G}^2$ terms may provide additional interactions that help stabilize the scalar field configuration under rapid rotation.


Notes that in both panels of Fig. \ref{fig:kappa_6 Qs}, all solutions exhibit two breakpoints on the left-hand side of the curves, with the exception of the $\alpha_2=0$ case, which actually exhibits three breakpoints: two corresponding to the connections between branch $1$ and branch $2$, and one at the right endpoint of branch $3$. Breakpoints are a common feature in scalar Gauss-Bonnet models \cite{PhysRevLett.112.251102,Kleihaus:2011tg,Kleihaus:2015aje,PhysRevD.90.124063,PhysRevD.54.5049,delgado2020spinning}. 
As black hole parameters approach the values of breakpoints, the numerical process fails to converge. This behavior arises because the quadratic equation governing the second-order near-horizon expansion of the scalar field ceases to yield real solutions near the breakpoints. Consequently, a uniform near-horizon expansion becomes infeasible, indicating the absence of a regular solution. For further details, refer to \cite{Kleihaus:2015aje,delgado2020spinning}.
\subsection{The results for \texorpdfstring{$\kappa$}{kappa} = 25}

\begin{figure}[htbp]
    \centering
    \begin{minipage}{0.32\textwidth}
    \centering
	\includegraphics[width=\textwidth]{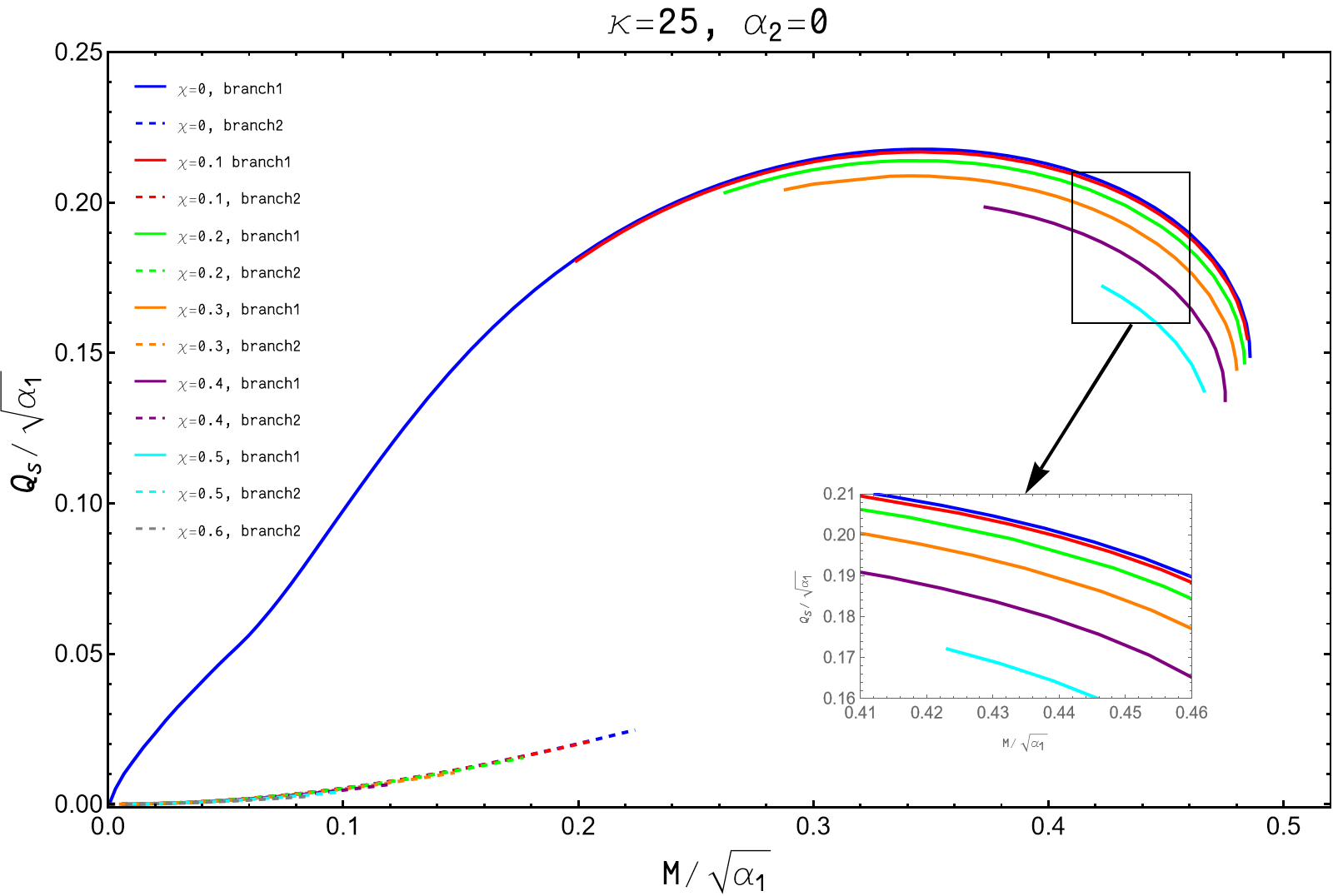}
    \end{minipage}
    \hspace{0.1cm}
    \begin{minipage}{0.32\textwidth}
	\centering
	\includegraphics[width=\textwidth]{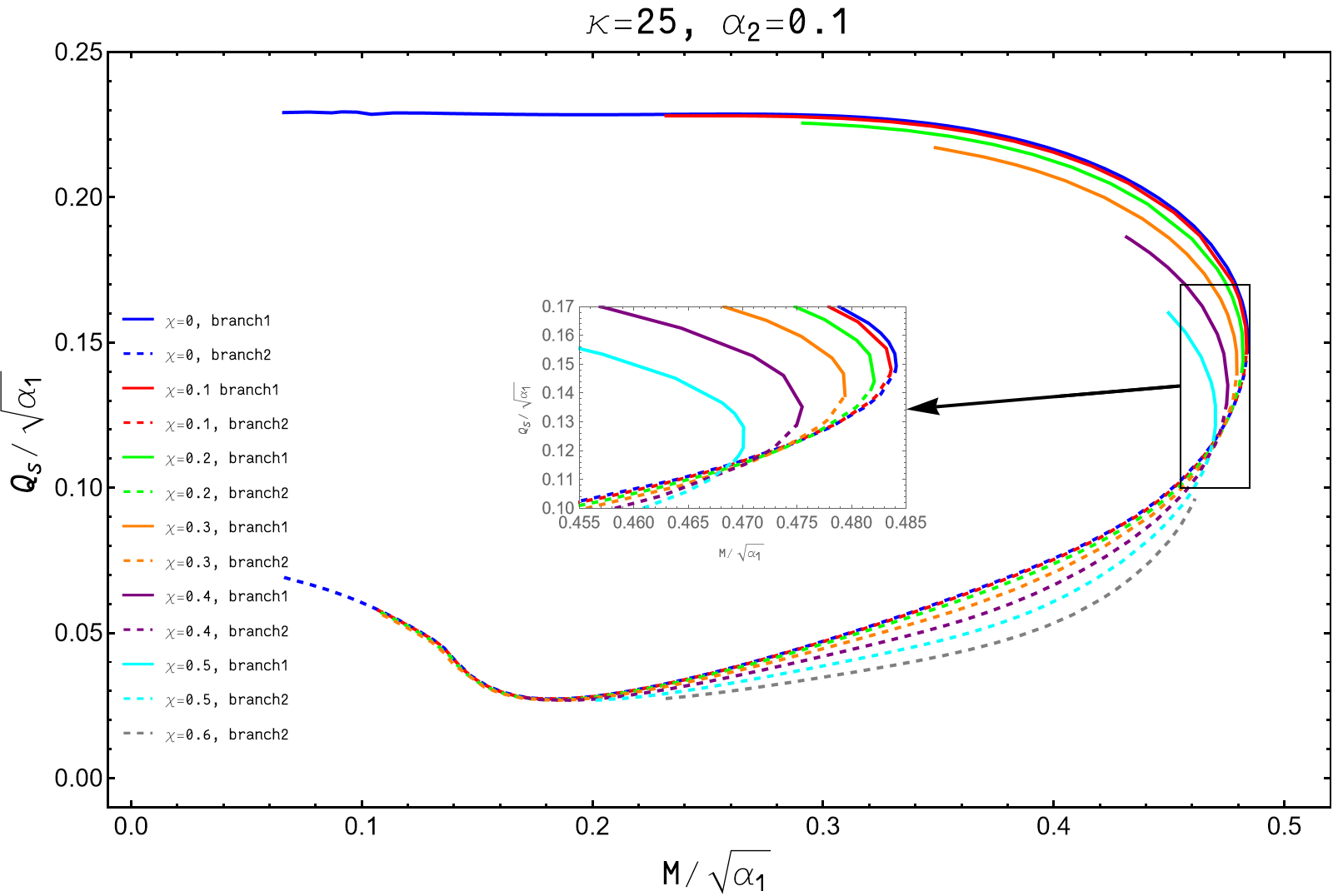}
    \end{minipage}
    \hspace{0.1cm}
    \begin{minipage}{0.32\textwidth}
	\centering
	\includegraphics[width=\textwidth]{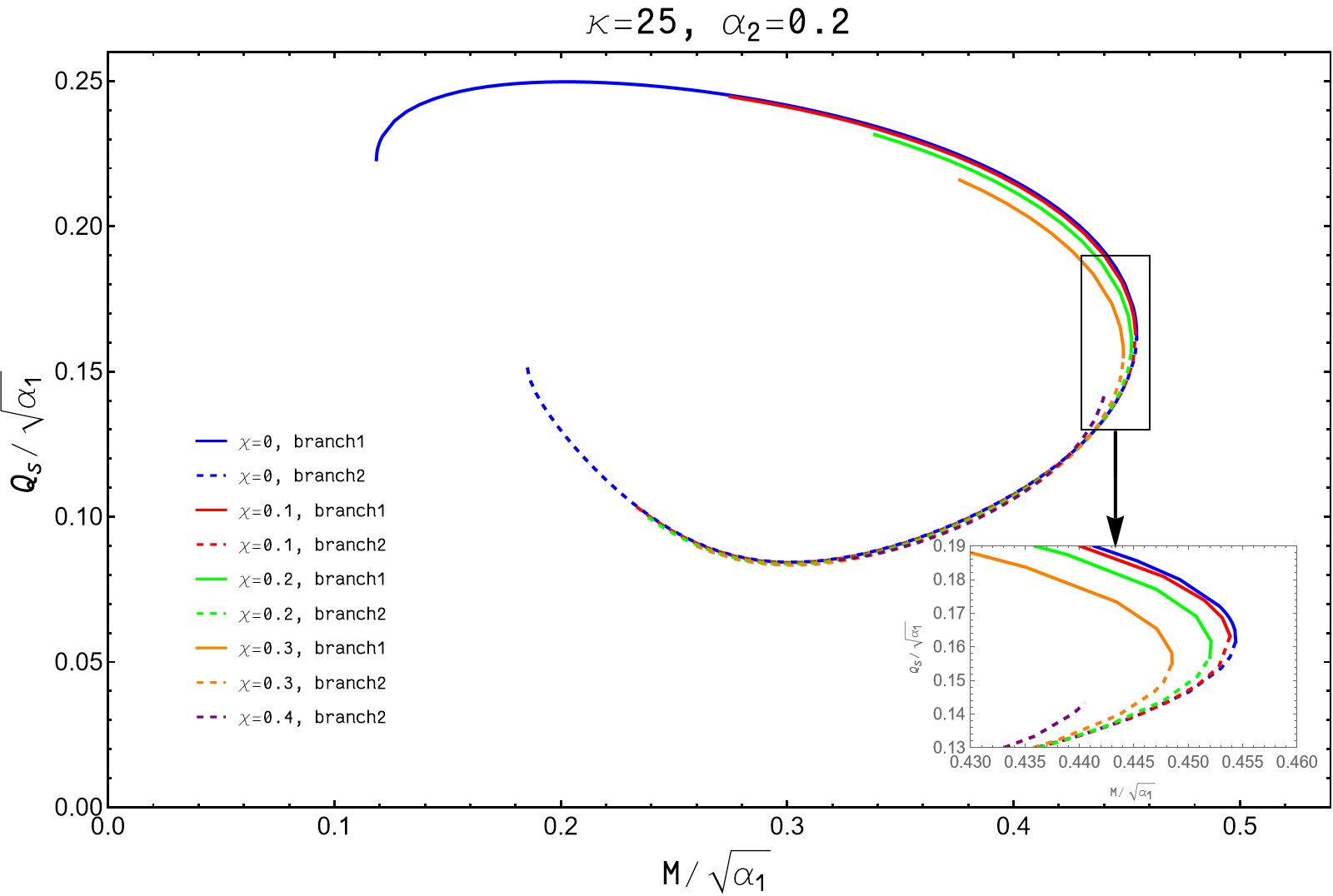}
    \end{minipage}
    \caption{The scalar charge $Q_s/\sqrt{\alpha_1}$ as a function of the ADM mass $M/\sqrt{\alpha_1}$ for $\kappa = 25$ when $\alpha_2 = 0$ (left), $\alpha_2 = 0.1$ (middle) and $\alpha_2 = 0.2$ (right). For each value of $\chi$, the solid lines 
    represent the upper branch, while the dashed lines represent the lower branch.}
    \label{fig:kappa_25 Qs}
\end{figure}
Considering $\kappa = 25$, we find that  no solutions exist when $\alpha_2>0.2$. 
We plot the scalar charge $Q_s$ as a function of the ADM mass $M$ for various values of the dimensionless spin parameter $\chi$ in Fig. \ref{fig:kappa_25 Qs}, with $\alpha_2 =0$, $0.1$, and $0.2$. 
Each spin value $\chi$ exhibits two branches: solid lines represent the upper branches and dashed lines represent the lower branches, with the same color used for each $\chi$.
The left panel of Fig.~\ref{fig:kappa_25 Qs} shows the scalar charge as a function of black hole mass for $\alpha_2 = 0$. 
The non-rotating ($\chi = 0$) scalarized black hole solutions share the same topological structure as those presented in \cite{Doneva:2021tvn,PhysRevD.105.L041502}. 
The results for $\alpha_2 = 0.1$ and $\alpha_2 = 0.2$ are presented in the middle and right panels of Fig. \ref{fig:kappa_25 Qs}, respectively.
For each rotating scalarized black holes with $\alpha_2 = 0.1$ and $\alpha_2 = 0.2$, there are two breakpoints located at the left endpoint of the solid and dashed lines of the same color. 
In contrast, for rotating scalarized black hole with $\alpha_2 = 0$, there are three breakpoints: two on the solid colored lines and one on the dashed line of the same color. 
By comparing the three panels, we observe that the domain of existence of the nonlinearly scalarized rotating black hole for $\kappa=25$ is significantly reduced when $\alpha_2$ increases from $0$ to $0.2$. 
To illustrate the effect more clearly, we provide the mass range of the existence domain $\Delta M$ for different $\chi$ in Table. \ref{tab:DeltaM_kappa_25}.
The value of $\Delta M$ decreases markedly as $\alpha_2$ rises from $0$ to $0.2$.
Moreover, by comparing $\Delta M$ for $\chi = 0$ with $\chi \neq 0$, we find the spin of nonlinearly scalarized rotating black holes also reduces the domain of existence.
\setlength{\tabcolsep}{4mm}  
\begin{table}[htb]
    \centering
    \caption{The values of $\Delta M$ for various $\chi$ with $\kappa = 25$ and $\alpha_2 = 0, 0.1, 0.2$.}
    \label{tab:DeltaM_kappa_25}
    \begin{tabular}{lcccccc}  
        \specialrule{1pt}{0pt}{0pt}  
        $\chi$           & 0 & 0.1 & 0.2 & 0.3 & 0.4 & 0.5  \\
        \specialrule{0.5pt}{0em}{0em}  
        $\alpha_2 = 0$   & 0.48576 & 0.48462 & 0.48343 & 0.48009 & 0.47519 & 0.46617  \\
        \specialrule{0.5pt}{0em}{0em}  
        $\alpha_2 = 0.1$ & 0.41794 & 0.37699 & 0.37394 & 0.37055 & 0.30490 & 0.26876  \\
        \specialrule{0.5pt}{0em}{0em}  
        $\alpha_2 = 0.2$ & 0.29680 & 0.22040 & 0.21366 & 0.18870 & 0.11777 & \multicolumn{1}{c}{---}  \\
        \specialrule{1pt}{0pt}{0pt}  
    \end{tabular}
\end{table}
\begin{figure}[htbp]
	 \begin{minipage}{0.5\textwidth}
        \centering
	\includegraphics[width=\textwidth]{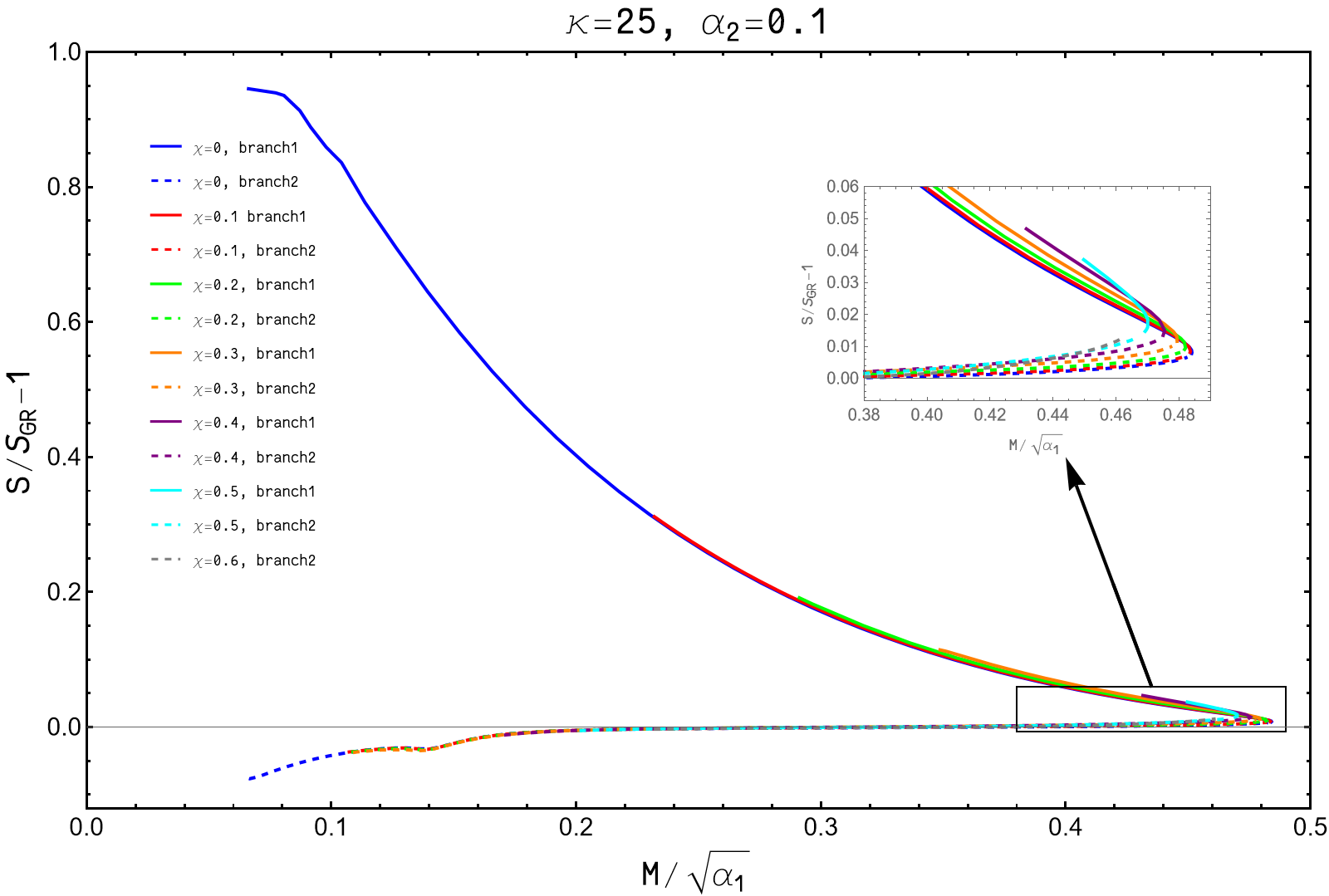}
  \end{minipage}
    \caption{Comparison of the entropy between scalarized black holes $S$ (with $\kappa=100$, $\alpha_2=0.1$) and Kerr black holes $S_{GR}$, plotted against $M/\sqrt{\alpha_1}$.
    }
    \label{fig:kappa_25 S}
\end{figure}

To explore the thermodynamic properties of black hole solutions, we compare the entropy $S$ of nonlinearly scalarized black holes with that of Kerr black holes $S_{GR}$ for various values of $\chi$, as shown in Fig. \ref{fig:kappa_25 S}. 
The entropy of the scalarized black holes on branch $1$ is always greater than that of both Kerr black holes and scalarized black holes on branch $2$, when compared at the same mass $M$ and spin $\chi$. 
In contrast, for most of the mass range, the entropy of black holes on branch $2$ is lower than that of the corresponding Kerr solution.
These results imply that within the mass range where branch 1 nonlinearly scalarized black holes exist, they are thermodynamically more stable than both Kerr black holes and branch 2 scalarized solutions with identical mass and spin. 

\begin{figure}[htbp]
    \centering
    \begin{minipage}{0.45\textwidth}
    \centering
	\includegraphics[width=\textwidth]{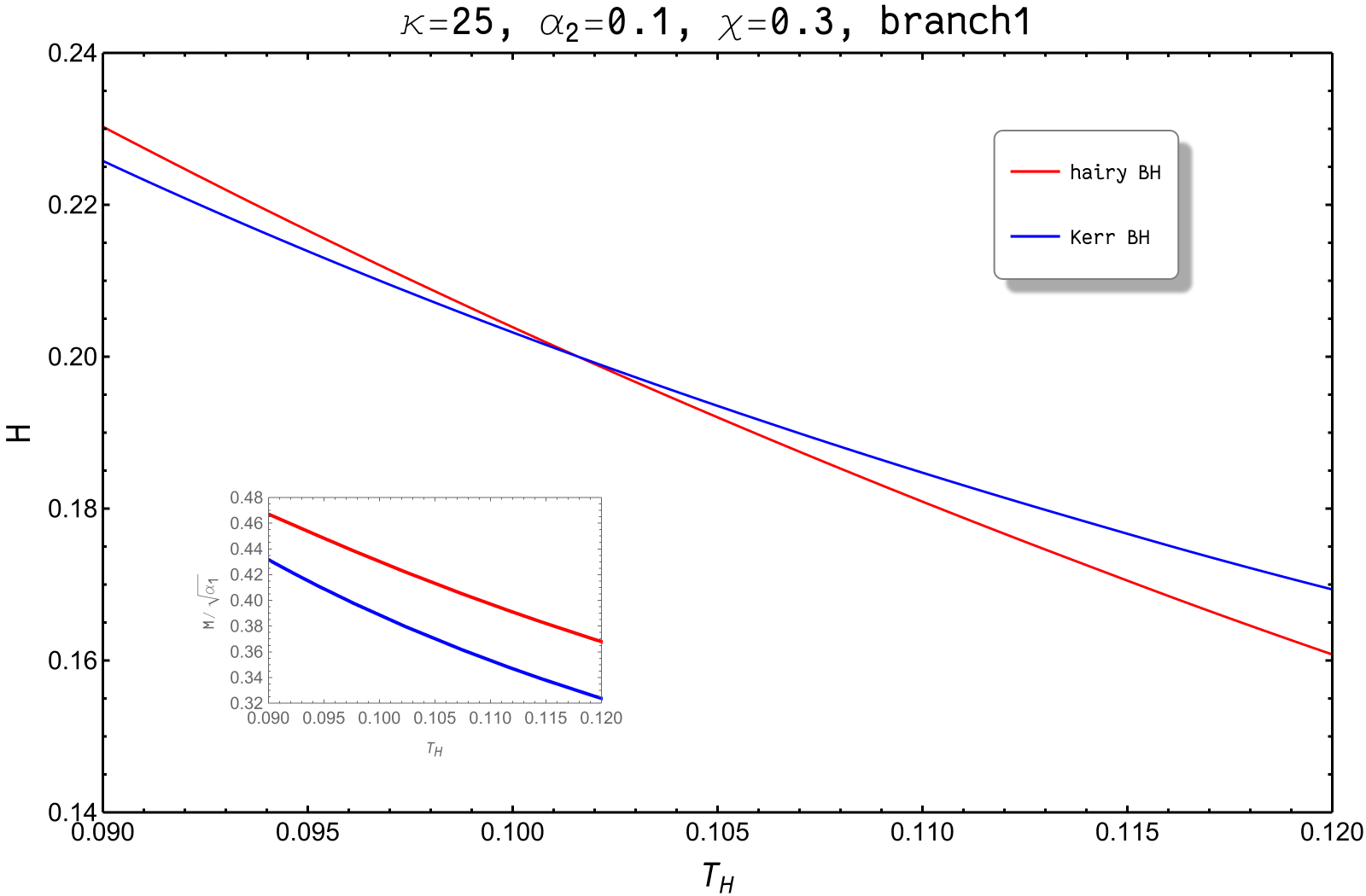}
    \end{minipage}
    \hspace{0.5cm}
    \begin{minipage}{0.45\textwidth}
	\centering
	\includegraphics[width=\textwidth]{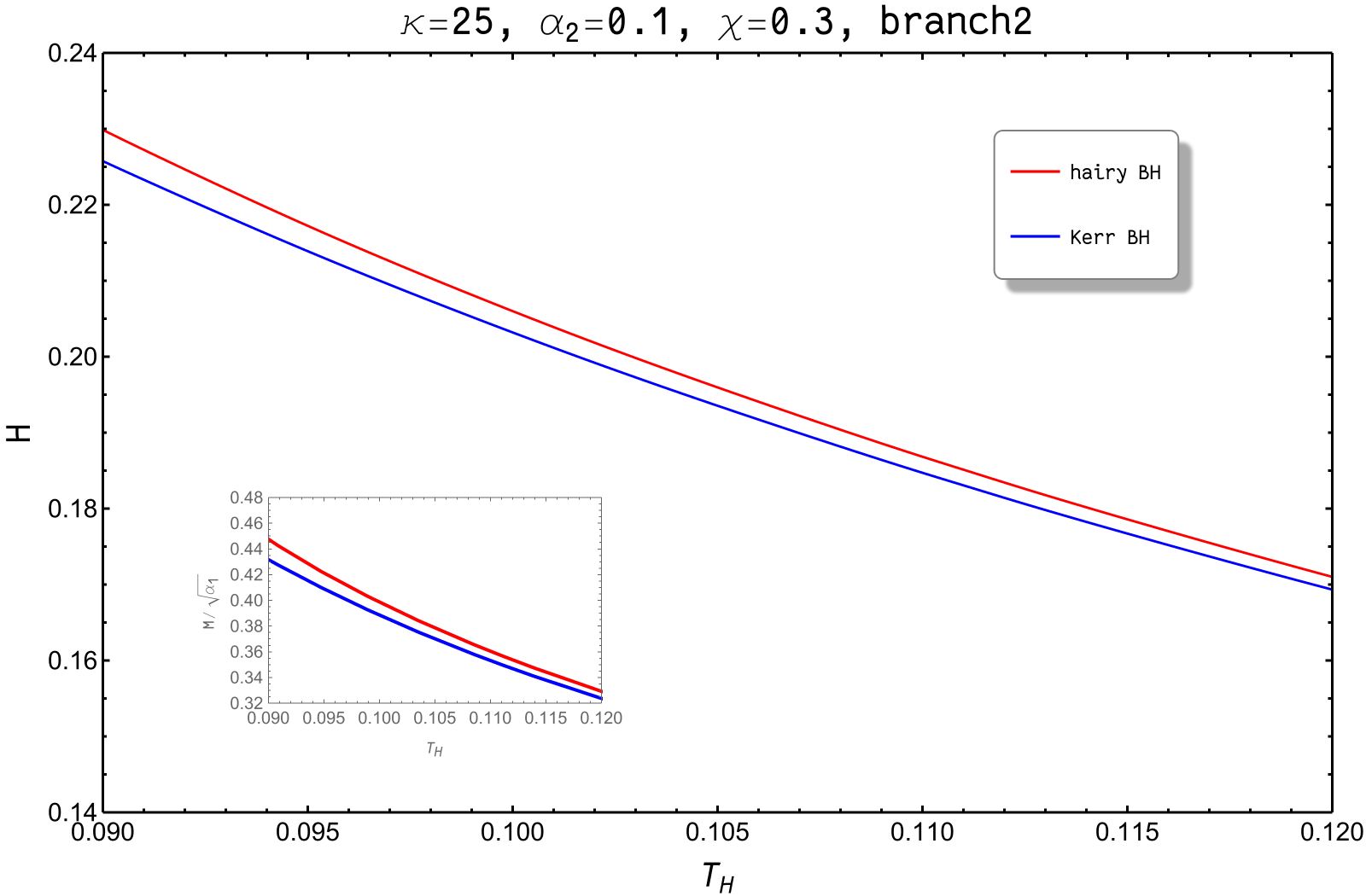}
    \end{minipage}
    \caption{Free energy $H$ as a function of Hawking temperature for Kerr and two branches of scalarized rotating black holes with $\kappa=25$, $\alpha_2=0.1$ and $\chi =0.3$.
    }
    \label{fig:kappa_25 H vs TH}
\end{figure}

Additionally, we examine the Helmholtz free energy (on-shell) $H=M-T_{H}S$ as a function of the Hawking temperature to investigate possible phase transitions between scalarized rotating black holes and Kerr black holes in canonical ensembles. 
For a fixed $\chi=0.3$, the free energy is depicted in Fig. \ref{fig:kappa_25 H vs TH}. 
The left panel shows that the free energy of the Kerr black hole is always lower than that of the scalarized rotating black hole for $T_H < T_c$. 
At the critical temperature $T_c \approx 0.10160$, the curves intersect, and for $T_H > T_c$, the free energy of the rotating scalarized  black hole becomes lower.
Thus, below $T_c$ the Kerr black hole is thermodynamically favored, while above $T_c$, the scalarized rotating black hole represents the ground state in the theory.
In comparison with branch $1$, branch $2$ black holes consistently exhibit higher free energy than Kerr black holes, indicating that the scalarized rotating black holes on branch $2$ are thermodynamically unstable relative to Kerr black holes. Furthermore, we find that the critical temperature $T_c$ increases with the spin parameter $\chi$, as summarized in Table \ref{tab: kappa_25 Tc}. 
In particular, for $\chi \geq 0.4$, the Hawking temperature remains below $T_c$ throughout the considered regime.
\setlength{\tabcolsep}{4mm}  
\begin{table}[htb]
    \centering
    \caption{The critical temperature $T_c$ at discrete values of the spin parameter $\chi$ for $\kappa=25$ and $\alpha_2=0.1$.}
    \label{tab: kappa_25 Tc}
    \begin{tabular}{lcccccc}  
        \specialrule{1pt}{0pt}{0pt}  
        $\chi$           & 0 & 0.1 & 0.2 & 0.3   \\
        \specialrule{0.5pt}{0em}{0em}  
        $T_c$            & 0.09866 & 0.09901 & 0.10003 & 0.10160  \\
        \specialrule{1pt}{0pt}{0pt}  
    \end{tabular}
\end{table}

\subsection{The results for \texorpdfstring{$\kappa$}{kappa} = 100}

\begin{figure}[htbp]
    \centering
    \begin{minipage}{0.45\textwidth}
    \centering
	\includegraphics[width=\textwidth]{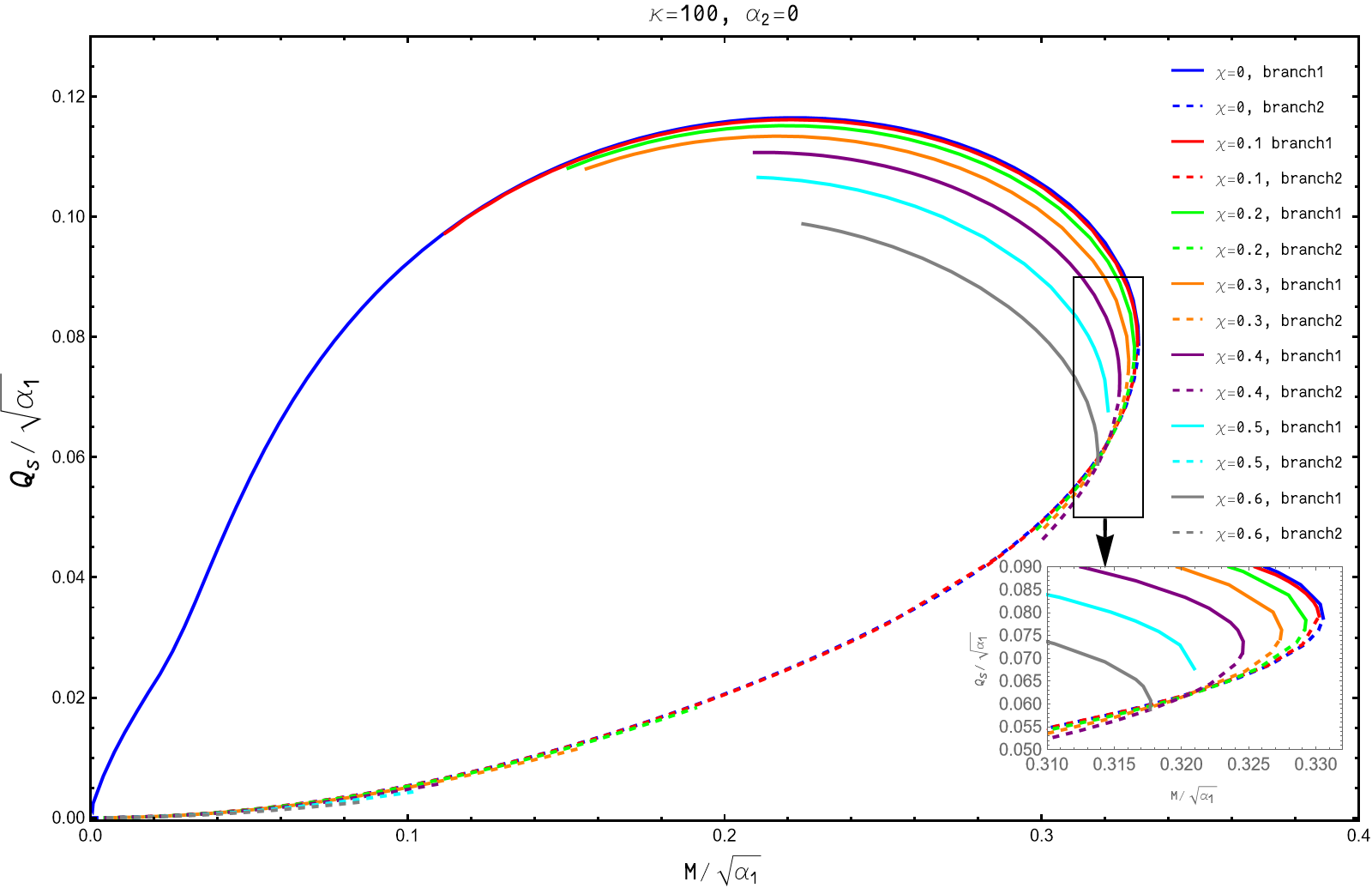}
    \end{minipage}
    \hspace{0.5cm}
    \begin{minipage}{0.45\textwidth}
	\centering
	\includegraphics[width=\textwidth]{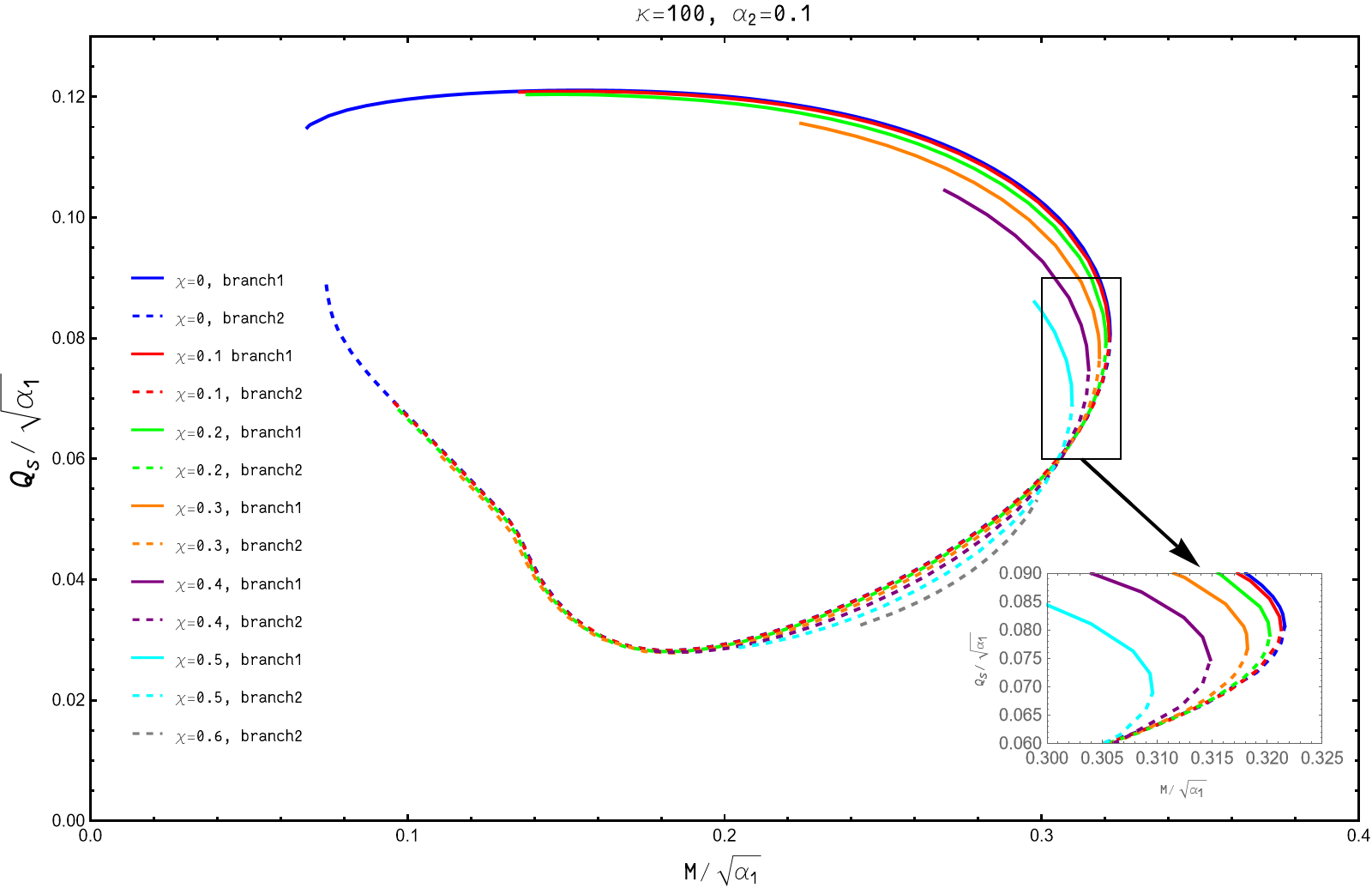}
    \end{minipage}
    \caption{The scalar charge $Q_s/\sqrt{\alpha_1}$ as a function of the ADM mass $M/\sqrt{\alpha_1}$ for $\kappa = 100$, $\alpha_2 = 0$ (left) and $\alpha_2 = 0.1$ (right). For each value of $\chi$, the solid lines 
    represent the upper branch, while the dashed lines represent the lower branch.}
    \label{fig:kappa_100 Qs}
\end{figure}

For the constant $\kappa = 100$, we obtained nonlinearly scalarized solutions with $\alpha_2 < 0.2$.We present the scalar charge $Q_s$ as a function of the ADM mass $M$ for various spin parameters $\chi$ in Fig. \ref{fig:kappa_100 Qs}. The left panel shows the result of $\alpha_2 = 0$ and the right panel shows the result of $\alpha_2 = 0.1$. Here, we also use solid lines to represent the upper branches and dashed lines to represent the lower branches, but the same color for the same $\chi$. Note that the result of non-rotating scalarized solutions have the same topological structure as Ref. \cite{PhysRevD.105.L041502} and the result of rotating scalarized solutions have the same topological structure as Ref. \cite{ZouPhysRevD.108.084007}. Comparing the two panels, we find that the domain of existence of the scalarized solutions is still suppressed by the additional term $\mathcal{G}^2$. 
We also list the quantitative results of the domain of existence $\Delta M = M_{max} - M_{min}$ for different $\chi$ and $\alpha_2$ in Table \ref{tab:DeltaM_kappa_100}. The parameter $\chi$ also suppresses the domain of existence of scalarized black holes. This is in contrast to the result obtained in the decoupling limit \cite{PhysRevD.106.104027}, which shows that the larger the rotation parameter $\chi$, the larger the domain of existence of the scalarized black holes. Comparing the result of $\kappa = 25$ (Fig. \ref{fig:kappa_25 Qs}) and $\kappa = 100$ (Fig. \ref{fig:kappa_100 Qs}), we find that weak coupling ($\kappa = 100$) has a smaller scalar charge and a smaller domain of existence than strong coupling ($\kappa = 25$) for the same $\chi$. This is different from the result of coupling function $F(\phi)= \frac{1}{\kappa}(1-e^{-\kappa \phi^{2}})$ \cite{liu2025rotatingblackholesclass}, which shows that the larger the coupling constant $\kappa$, the larger the scalar charge and the domain of existence of the scalarized black holes for the same $\chi$.
\setlength{\tabcolsep}{4mm}  
\begin{table}[htb]
    \centering
    \caption{The values of $\Delta M $ for various $\chi$ with $\kappa = 100$ and $\alpha_2 = 0, 0.1$.}
    \label{tab:DeltaM_kappa_100}
    \begin{tabular}{lcccccc}  
        \specialrule{1pt}{0pt}{0pt}  
        $\chi$           & 0 & 0.1 & 0.2 & 0.3 & 0.4 & 0.5  \\
        \specialrule{0.5pt}{0em}{0em}  
        $\alpha_2 = 0$   & 0.33014 & 0.32582 & 0.32517 & 0.32116 & 0.32023 & 0.31587  \\
        \specialrule{0.5pt}{0em}{0em}  
        $\alpha_2 = 0.1$ & 0.25324 & 0.22566 & 0.22320 & 0.20772 & 0.13708 & 0.10522  \\
        \specialrule{1pt}{0pt}{0pt}  
    \end{tabular}
\end{table}

\begin{figure}[htbp]
	 \begin{minipage}{0.5\textwidth}
        \centering
	\includegraphics[width=\textwidth]{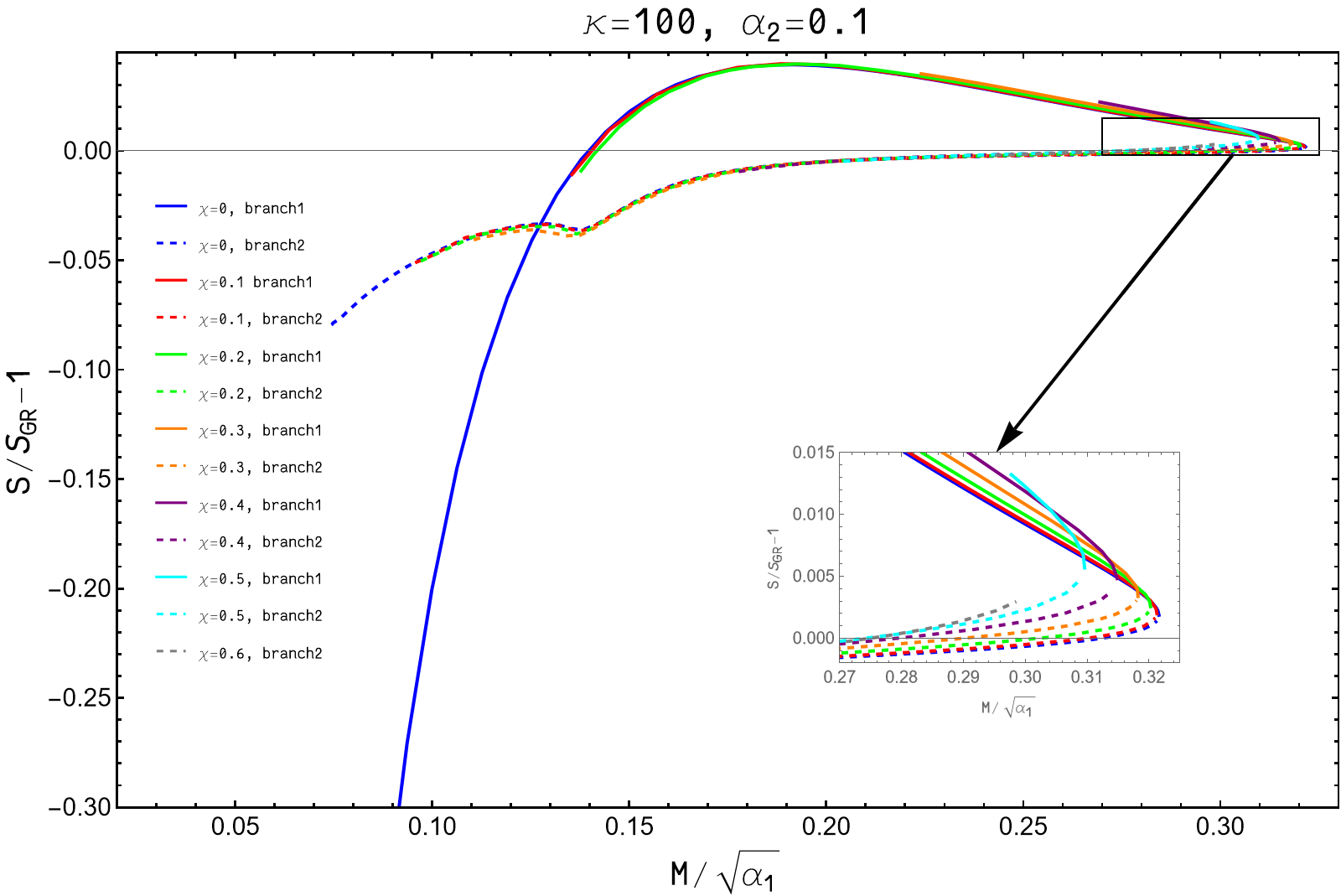}
  \end{minipage}
    \caption{Entropy comparison between scalarized black holes ($\kappa=100$, $\alpha_2=0.1$) and Kerr black holes $S_{GR}$ as a function of $M/\sqrt{\alpha_1}$.
    }
    \label{fig:kappa_100 S}
\end{figure}

\begin{figure}[htbp]
    \centering
    \begin{minipage}{0.45\textwidth}
    \centering
	\includegraphics[width=\textwidth]{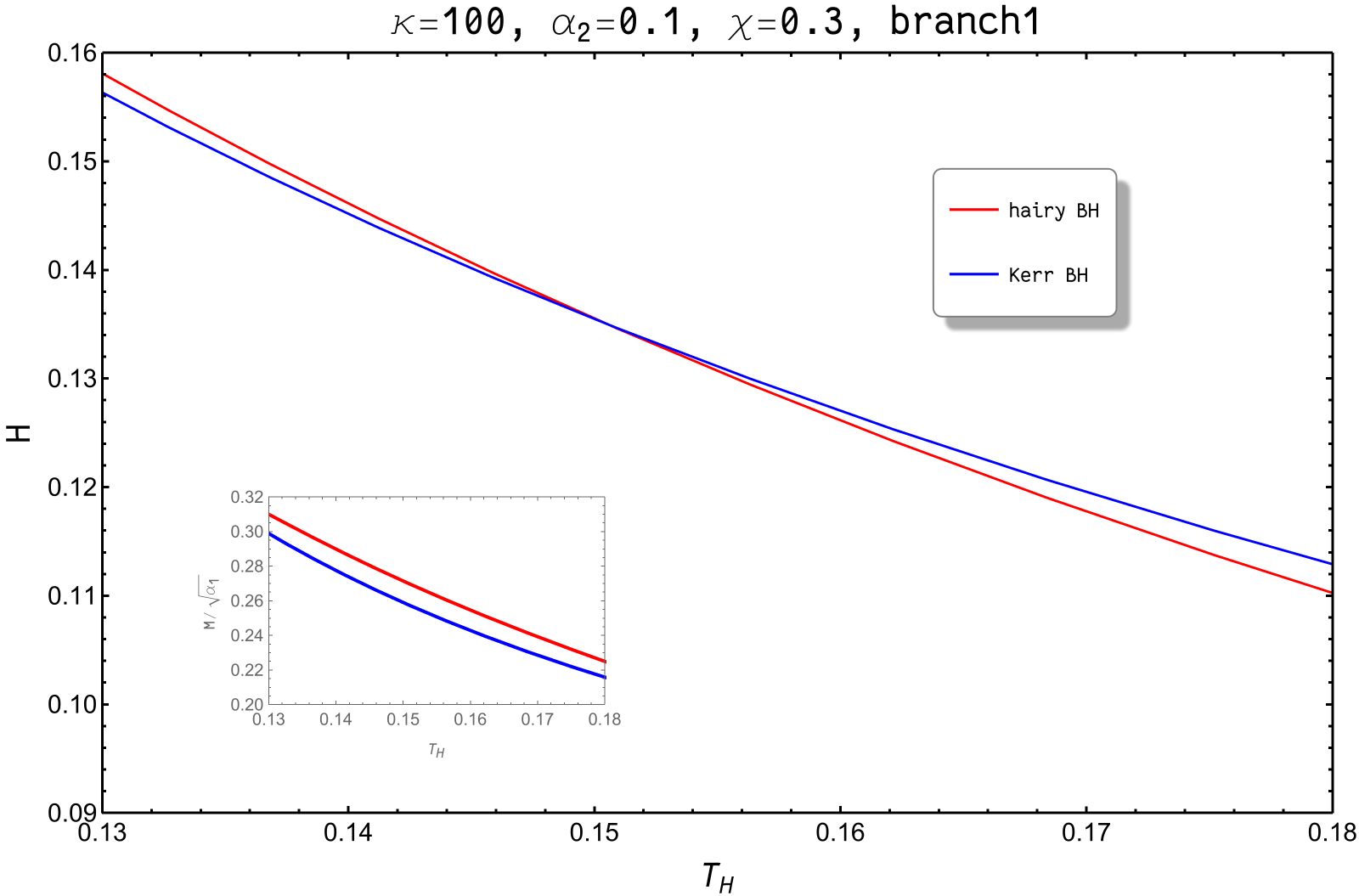}
    \end{minipage}
    \hspace{0.5cm}
    \begin{minipage}{0.45\textwidth}
	\centering
	\includegraphics[width=\textwidth]{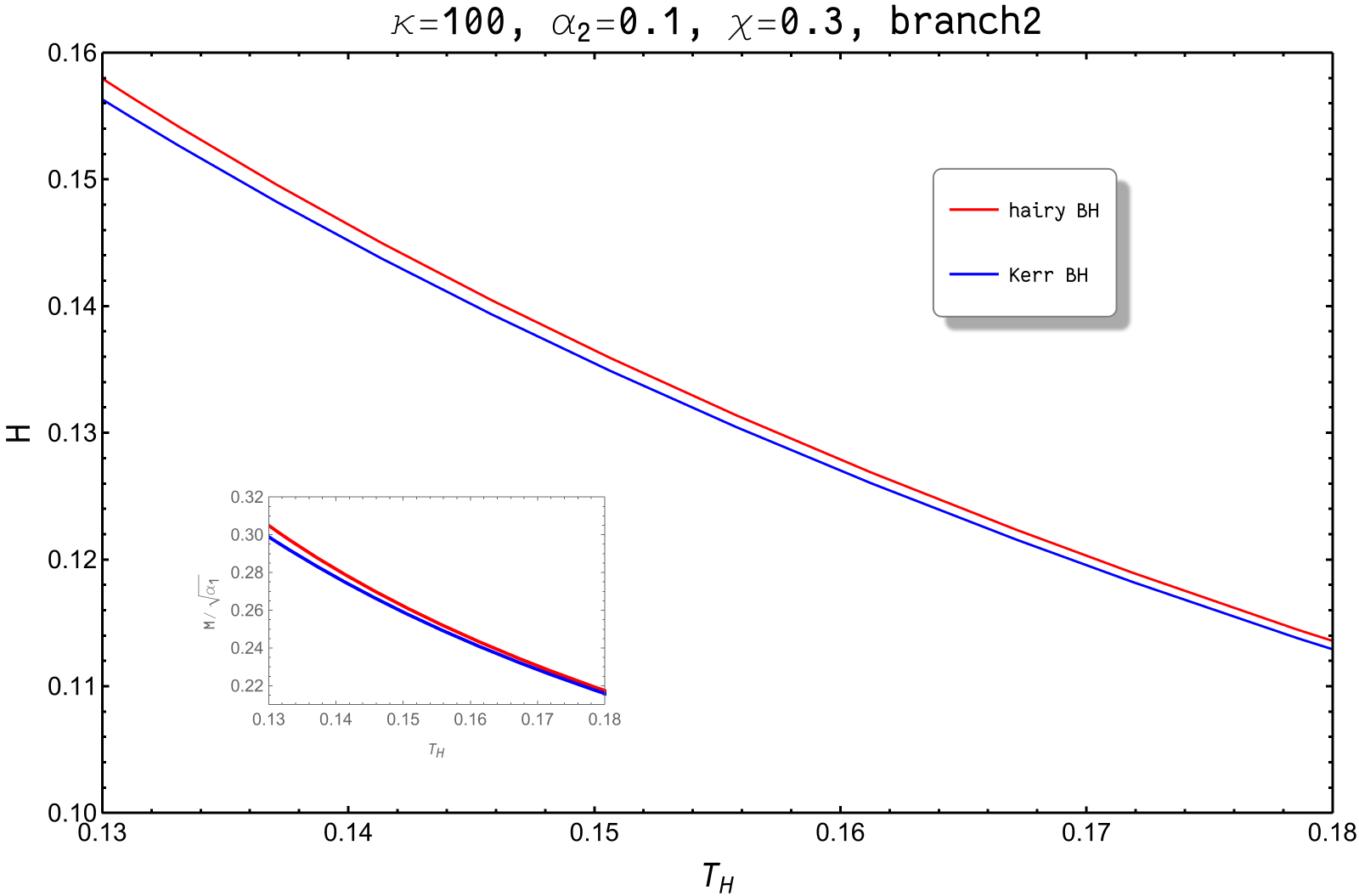}
    \end{minipage}
    \caption{Free energy $H$ as a function of Hawking temperature for Kerr and two branches of scalarized rotating black holes with $\kappa=100$, $\alpha_2=0.1$ and $\chi =0.3$.
    }
    \label{fig:kappa_100 H vs TH}
\end{figure}

The thermodynamic properties of the scalarized black holes for $\kappa = 100$ and $\alpha_2 = 0.1$ are presented in Fig. \ref{fig:kappa_100 S} and Fig. \ref{fig:kappa_100 H vs TH}. 
Figure. \ref{fig:kappa_100 S} compares the entropy of scalarized black holes $S$ with that of Kerr black holes $S_{GR}$ for identical mass $M$ and spin $\chi$. 
Unlike the result of $\kappa = 25$ case, the entropy of branch $1$ scalarized black holes is lower than that of Kerr black holes at smaller mass, but becomes higher as the mass increases. 
In branch $2$, the behavior resembles the $\kappa = 25$ scenario: the entropy of the scalarized solutions remains lower than that of Kerr black holes across most of the mass range.
These results suggest that the scalarized black holes are entropically favored over Kerr black holes with the same $\chi$ and $M$ at larger masses in branch $1$. 

Meanwhile, Fig. \ref{fig:kappa_100 H vs TH} presents the free energy $H$ as a function of the Hawking temperature for both Kerr and scalarized rotating black holes across two branches. The critical temperature is approximately $T_c \approx 0.15059$. 
For $T_H < T_c$, the ground state corresponds to the Kerr black hole, whereas for $T_H > T_c$, the scalarized rotating black hole becomes thermodynamically preferred. 
This indicates a possible first-order phase transition between Kerr and scalarized rotating black holes. In contrast, on branch $2$, the free energy of the scalarized black holes are always higher than Kerr black holes, implying that these solutions are thermodynamically less stable across the entire temperature range. 
Furthermore, as observed in the cases $\kappa=25$, $\alpha_2=0.1$ and $\chi=0.3$, the critical temperature $T_c$ increases with the spin parameter $\chi$, as summarized in Table. \ref{tab: kappa_100 Tc}. For values of $\chi \geq 0.4$, the temperature of the black hole is always less than the critical temperature.
\setlength{\tabcolsep}{4mm}  
\begin{table}[htb]
    \centering
    \caption{The critical temperature $T_c$ at discrete values of the spin parameter $\chi$ for $\kappa=100$ and $\alpha_2=0.1$.}
    \label{tab: kappa_100 Tc}
    \begin{tabular}{lcccccc}  
        \specialrule{1pt}{0pt}{0pt}  
        $\chi$           & 0 & 0.1 & 0.2 & 0.3   \\
        \specialrule{0.5pt}{0em}{0em}  
        $T_c$            & 0.14643 & 0.14693 & 0.14839 & 0.15059  \\
        \specialrule{1pt}{0pt}{0pt}  
    \end{tabular}
\end{table}

\begin{figure}[htbp]
    \centering
    \begin{minipage}{0.45\textwidth}
        \centering
	\includegraphics[width=\textwidth]{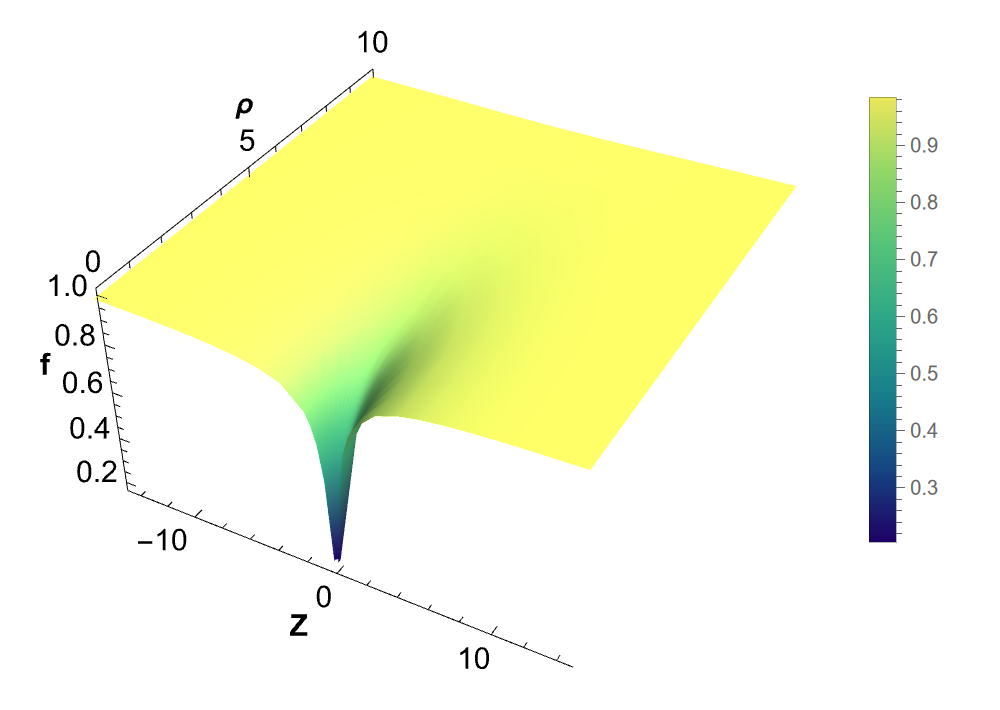}
    \end{minipage}
    \hspace{0.5cm}
    \begin{minipage}{0.45\textwidth}
	\centering
	\includegraphics[width=\textwidth]{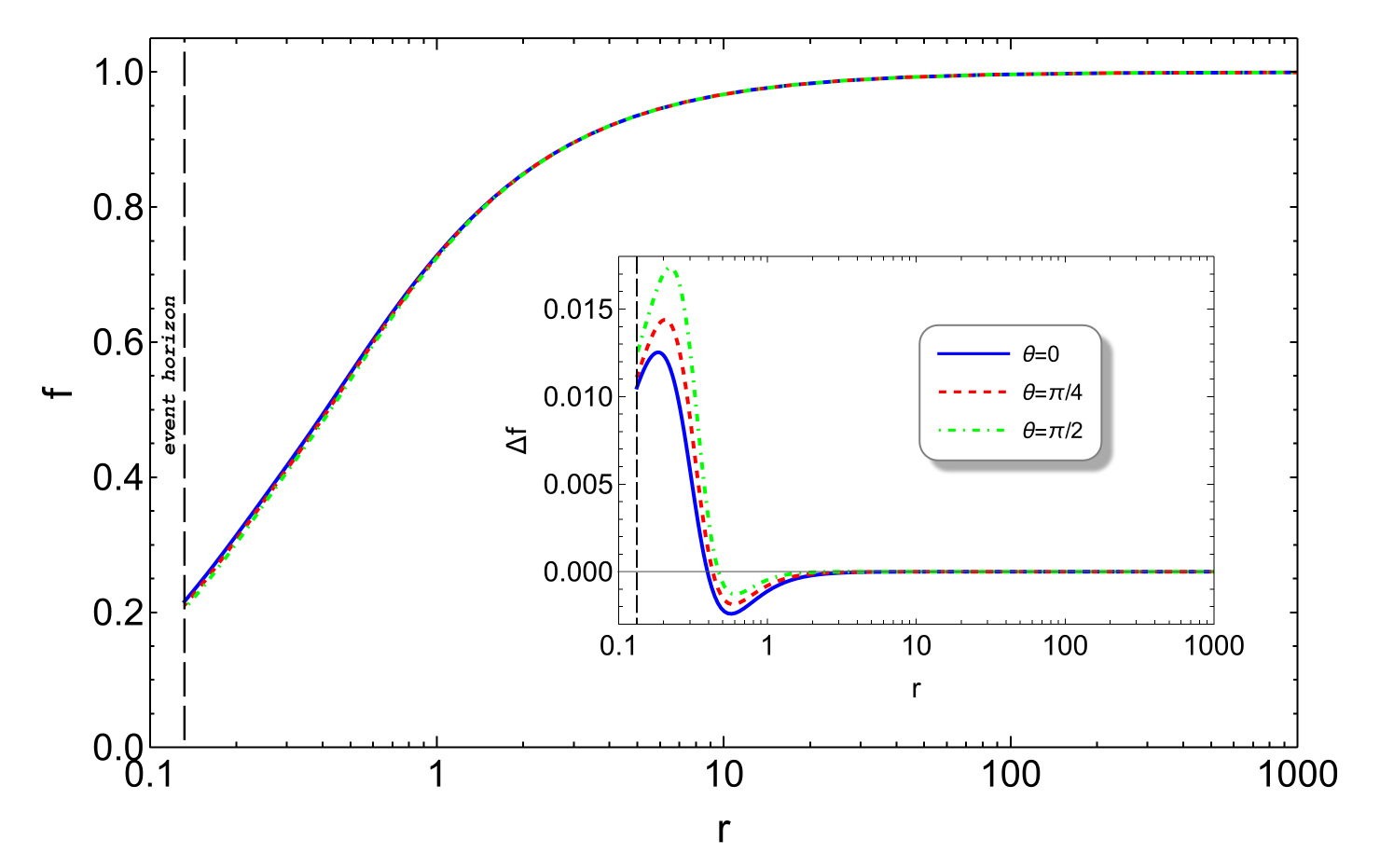}
    \end{minipage}
    
    \begin{minipage}{0.45\textwidth}
        \centering
	\includegraphics[width=\textwidth]{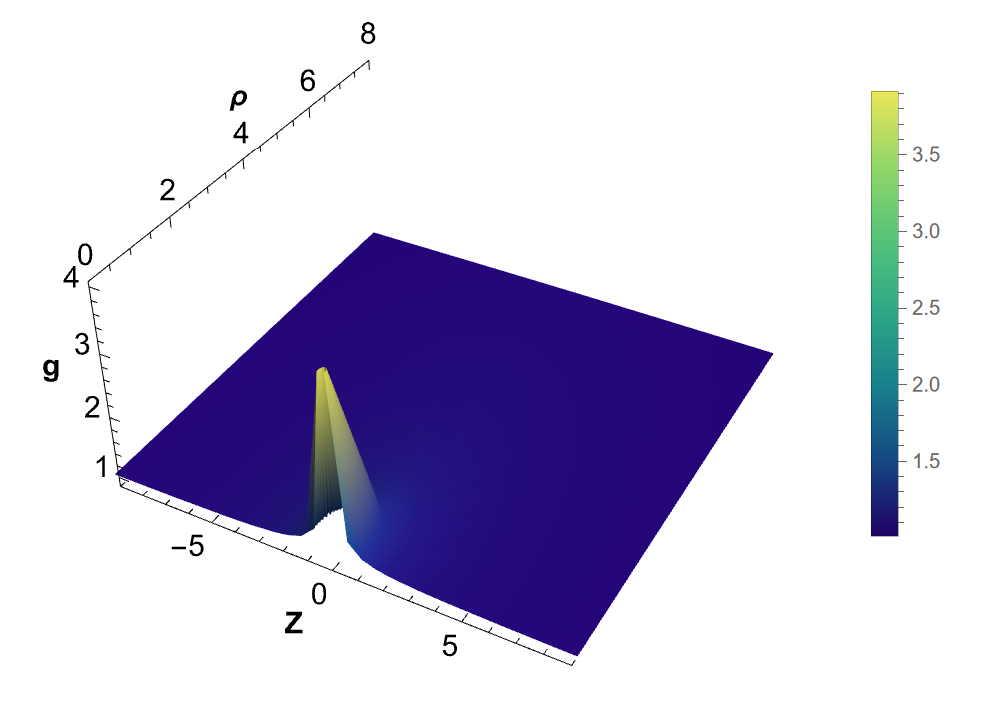}
    \end{minipage}
    \hspace{0.5cm}
    \begin{minipage}{0.45\textwidth}
	\centering
	\includegraphics[width=\textwidth]{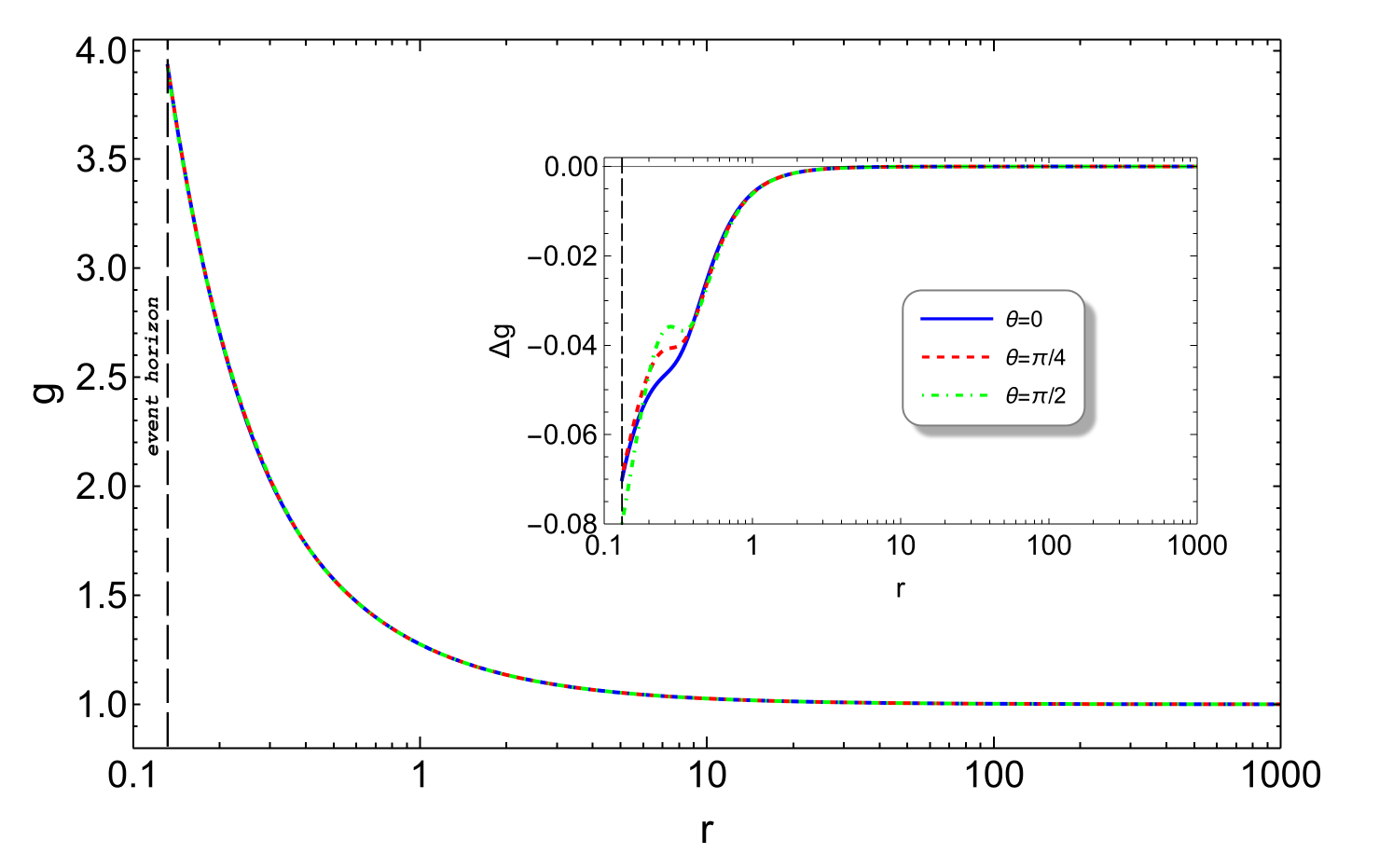}
    \end{minipage}
    
    \caption{Metric functions $f$ and $g$ for scalarized rotating black hole solution with the parameters $\chi = 0.4$, $r_H = 0.132$ (dashed black line) and $\alpha_2$ = $0.1$. The deviations between the scalarized black hole and the Kerr black hole are described by $\Delta f = f - f_{Kerr}$ and $\Delta g = g - g_{Kerr}$.}
    \label{fig:3D 2D metric functions}
\end{figure}

\begin{figure}[htbp]
    \centering
    \begin{minipage}{0.45\textwidth}
        \centering
	\includegraphics[width=\textwidth]{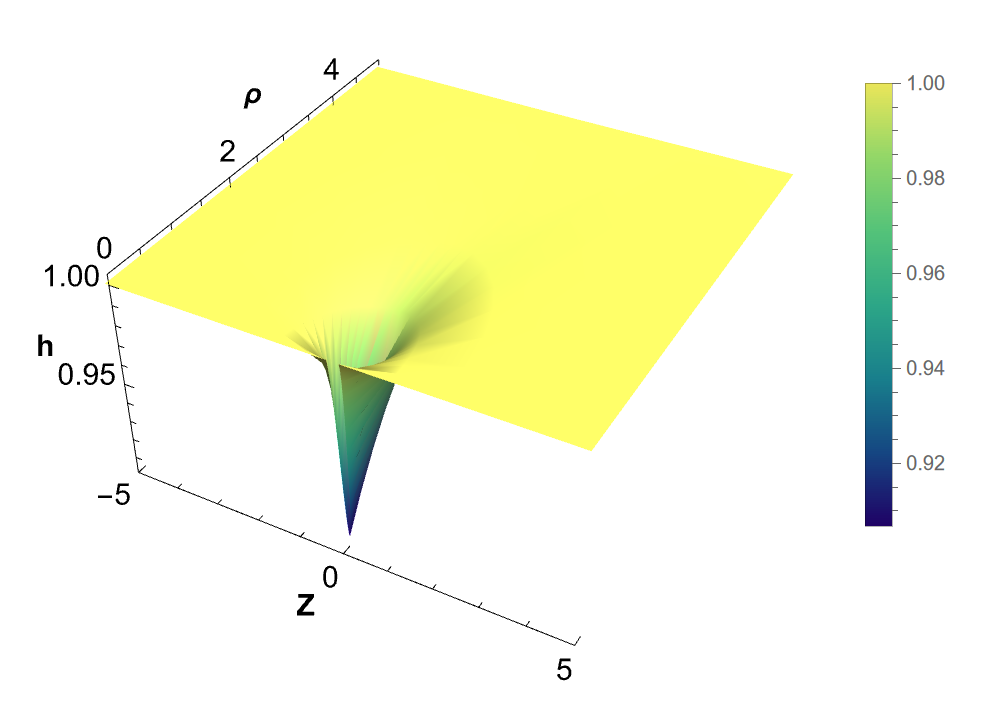}
    \end{minipage}
    \hspace{0.5cm}
    \begin{minipage}{0.45\textwidth}
	\centering
	\includegraphics[width=\textwidth]{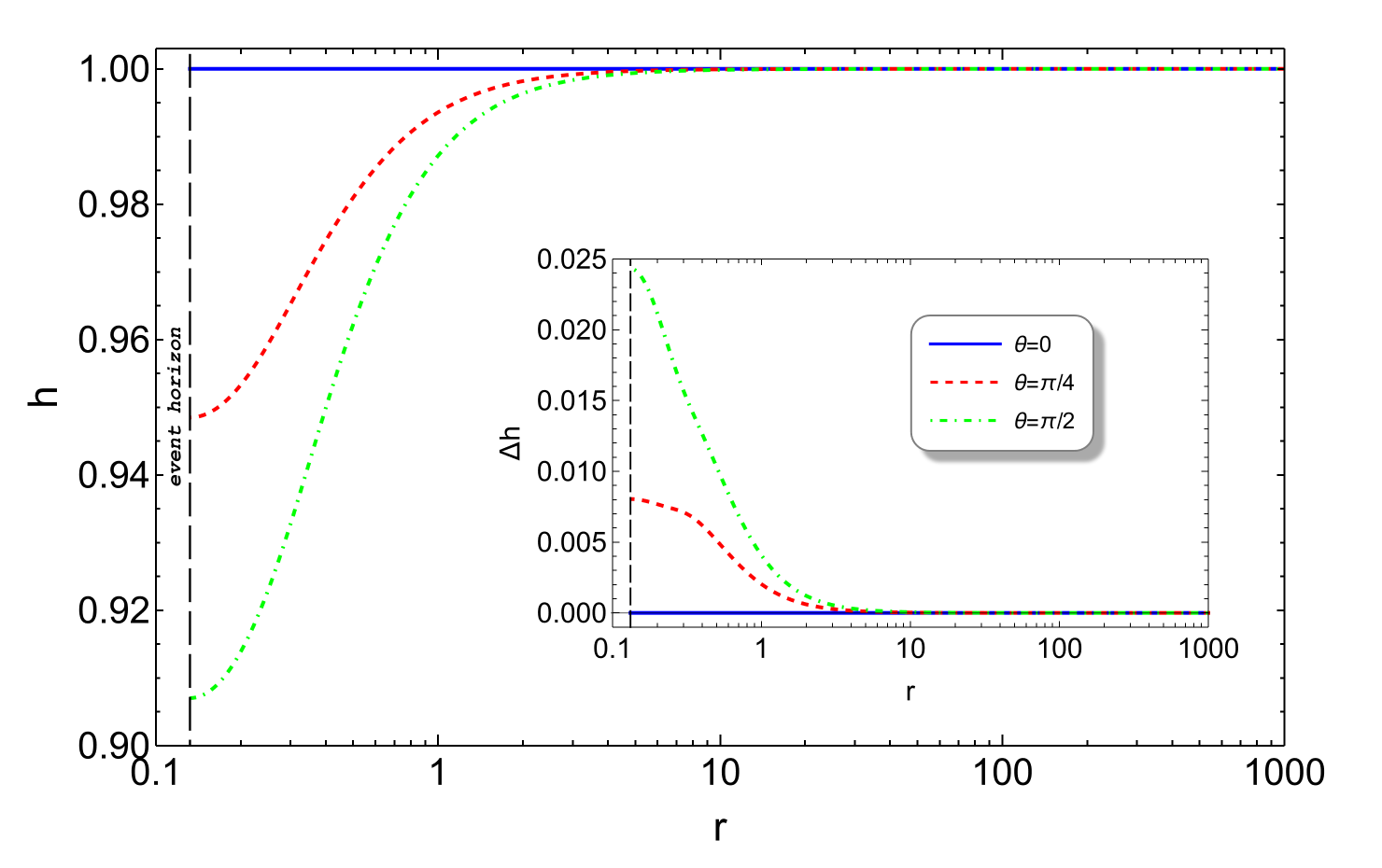}
    \end{minipage}
    
    \begin{minipage}{0.45\textwidth}
        \centering
	\includegraphics[width=\textwidth]{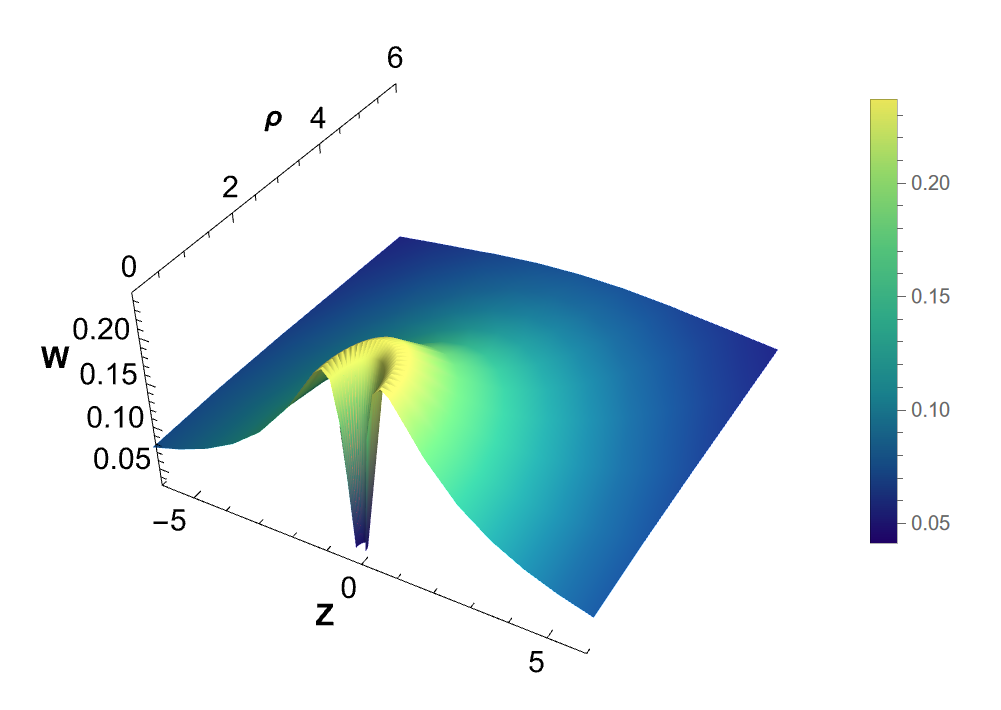}
    \end{minipage}
    \hspace{0.5cm}
    \begin{minipage}{0.45\textwidth}
	\centering
	\includegraphics[width=\textwidth]{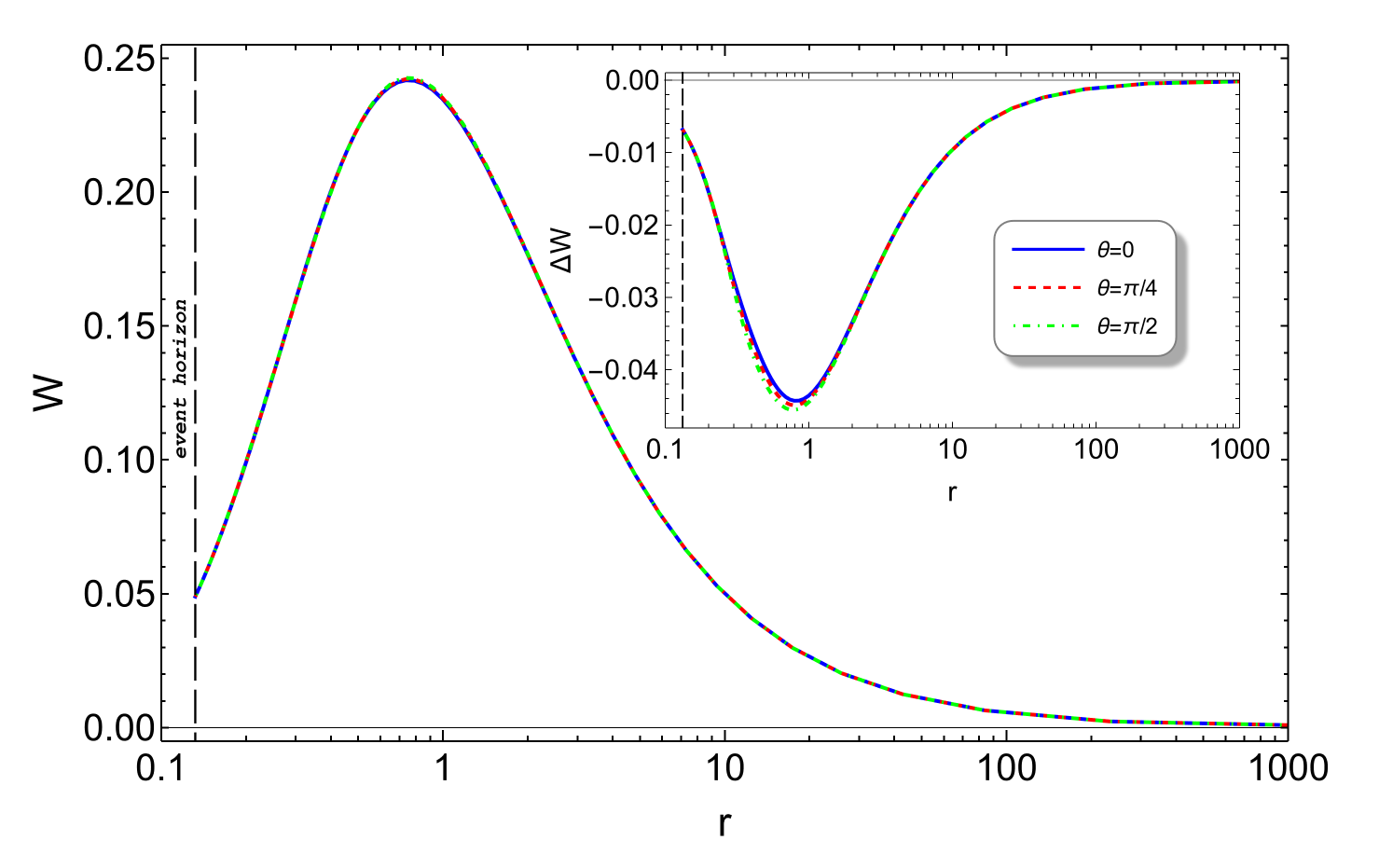}
    \end{minipage}

    \begin{minipage}{0.45\textwidth}
        \centering
	\includegraphics[width=\textwidth]{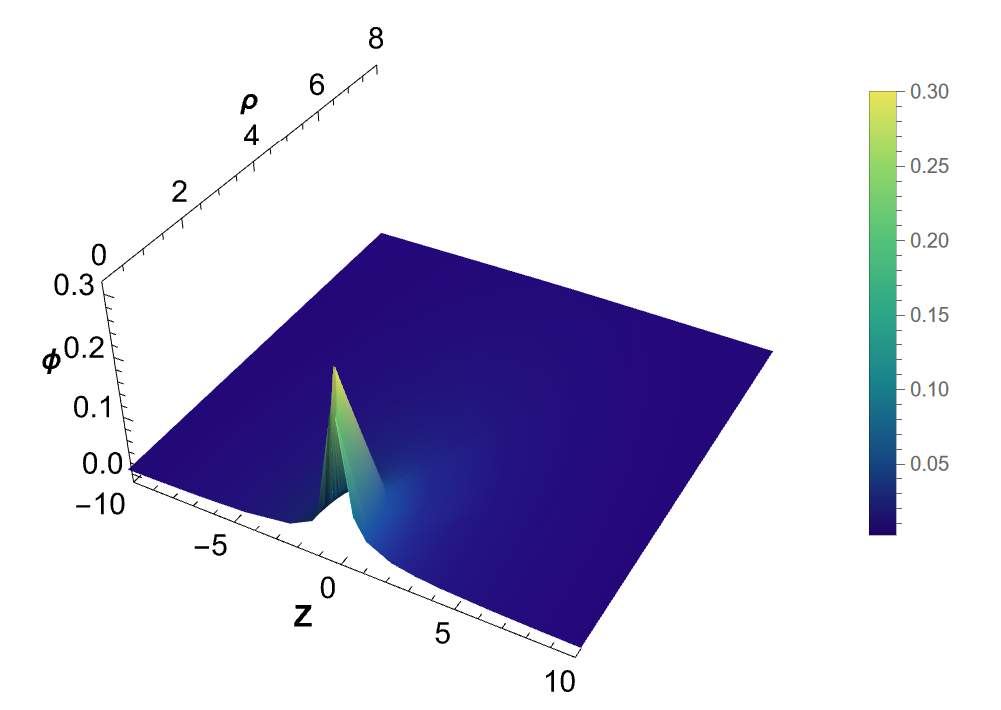}
    \end{minipage}
    \hspace{0.5cm}
    \begin{minipage}{0.45\textwidth}
	\centering
	\includegraphics[width=\textwidth]{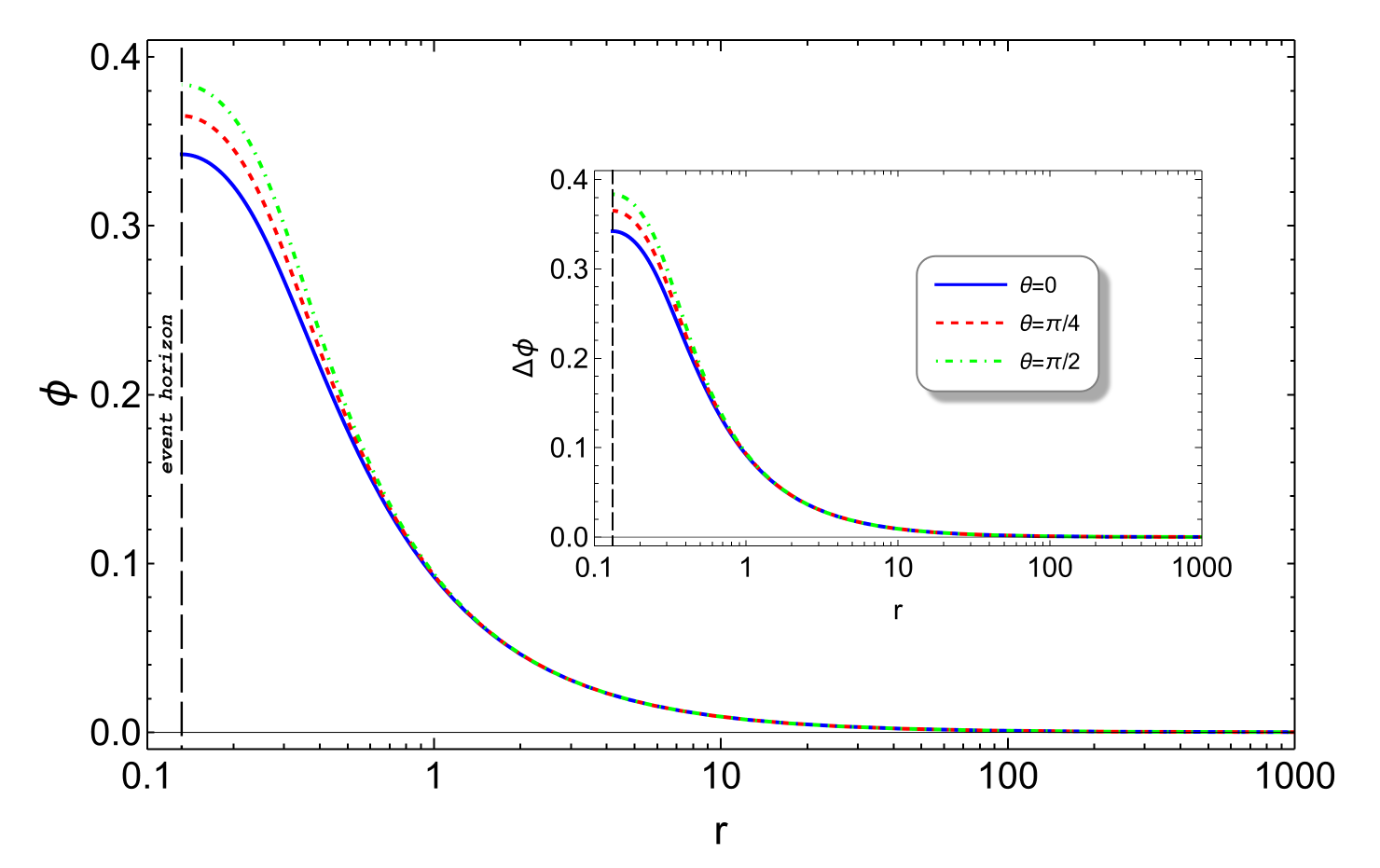}
    \end{minipage}

    \begin{minipage}{0.45\textwidth}
        \centering
	\includegraphics[width=\textwidth]{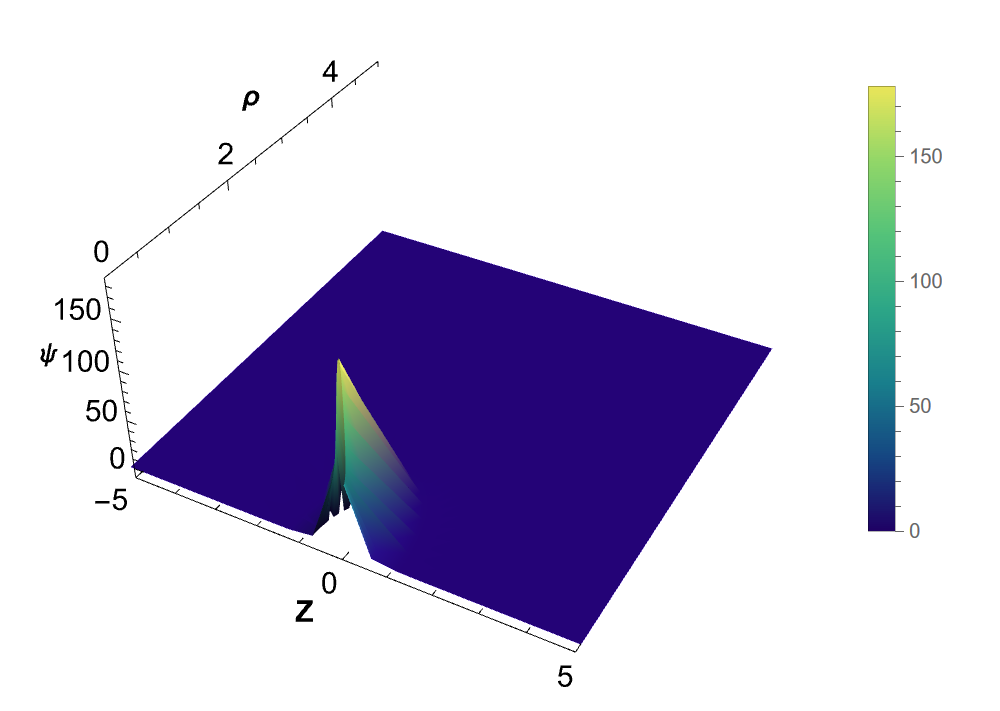}
    \end{minipage}
    \hspace{0.5cm}
    \begin{minipage}{0.45\textwidth}
	\centering
	\includegraphics[width=\textwidth]{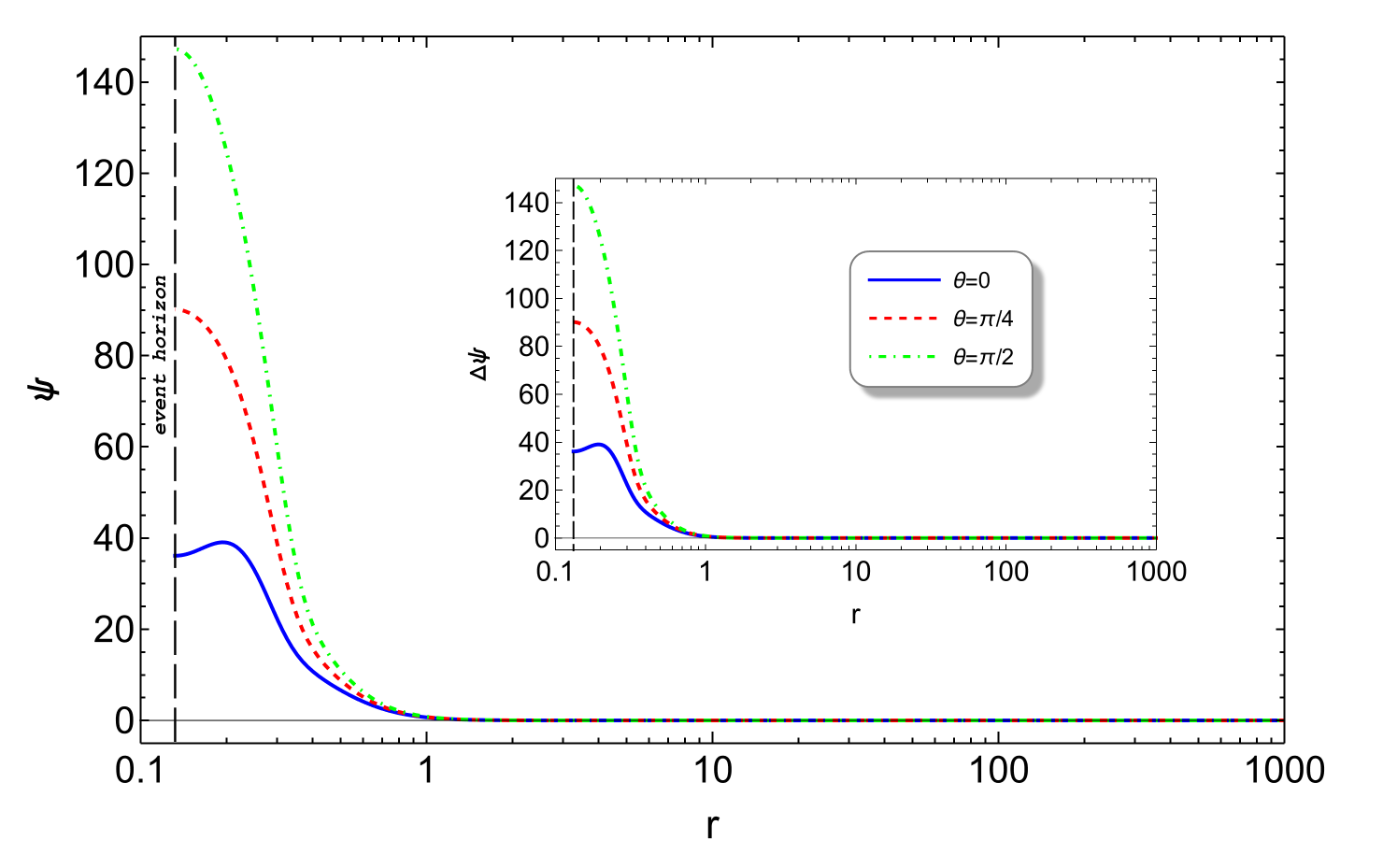}
    \end{minipage}
    \caption{Metric functions $h$, $W$ and scalar fields for scalarized rotating black hole solution with the same parameters as in Fig. \ref{fig:3D 2D metric functions}. The deviations between the scalarized black hole and the Kerr black hole are described by $\Delta h = h - h_{Kerr}$ and $\Delta W = W - W_{Kerr}$ .}
    \label{fig:3D 2D scalar fields}
\end{figure}

The metric functions and scalar fields $\mathcal{F}^{k}$ (with $\mathcal{F}^{k} = \{f, g, h, W, \phi, \psi \}$) and their first and second derivatives with respect to $r$ and $\theta$ have smooth profiles, which leads to finite curvature on the full domain of integration, specially on the event horizon. Moreover, the shape of $\mathcal{F}^{k}$ is rather similar to the Kerr black holes. To illustrate these features, we present the profile functions of a typical solution together with the deviations between the scalarized rotating black hole and the Kerr black hole in Figs. \ref{fig:3D 2D metric functions} and \ref{fig:3D 2D scalar fields}. Setting $\kappa=100$ and $\alpha_2 =0.1$, and the parameters $\chi = 0.4$, $r_H = 0.132$, we obtained the numerical solution with the black hole mass $M = 0.30032$ and the scalar charge $Q_s = 0.09268$. In these figures, the left columns show three-dimensional ($3$D) plots, and the right columns display two-dimensional ($2$D) plots of metric functions and scalar fields as a function of radial variable for different angular values. In the $3$D figures, the axes are $\rho = r \sin \theta$ and $z = r \cos \theta$ (with $r > r_H$). One can see from it that the numerical solution exhibit smooth profiles. As shown in $2$D plots, the Gauss-Bonnet term (recall that Eq. \ref{LM}, $\psi = \mathcal{G}$) stay finite everywhere, in particular at the horizon.

\begin{figure}[htbp]
    \centering
    \begin{minipage}{0.4\textwidth}
        \centering
	\includegraphics[width=\textwidth]{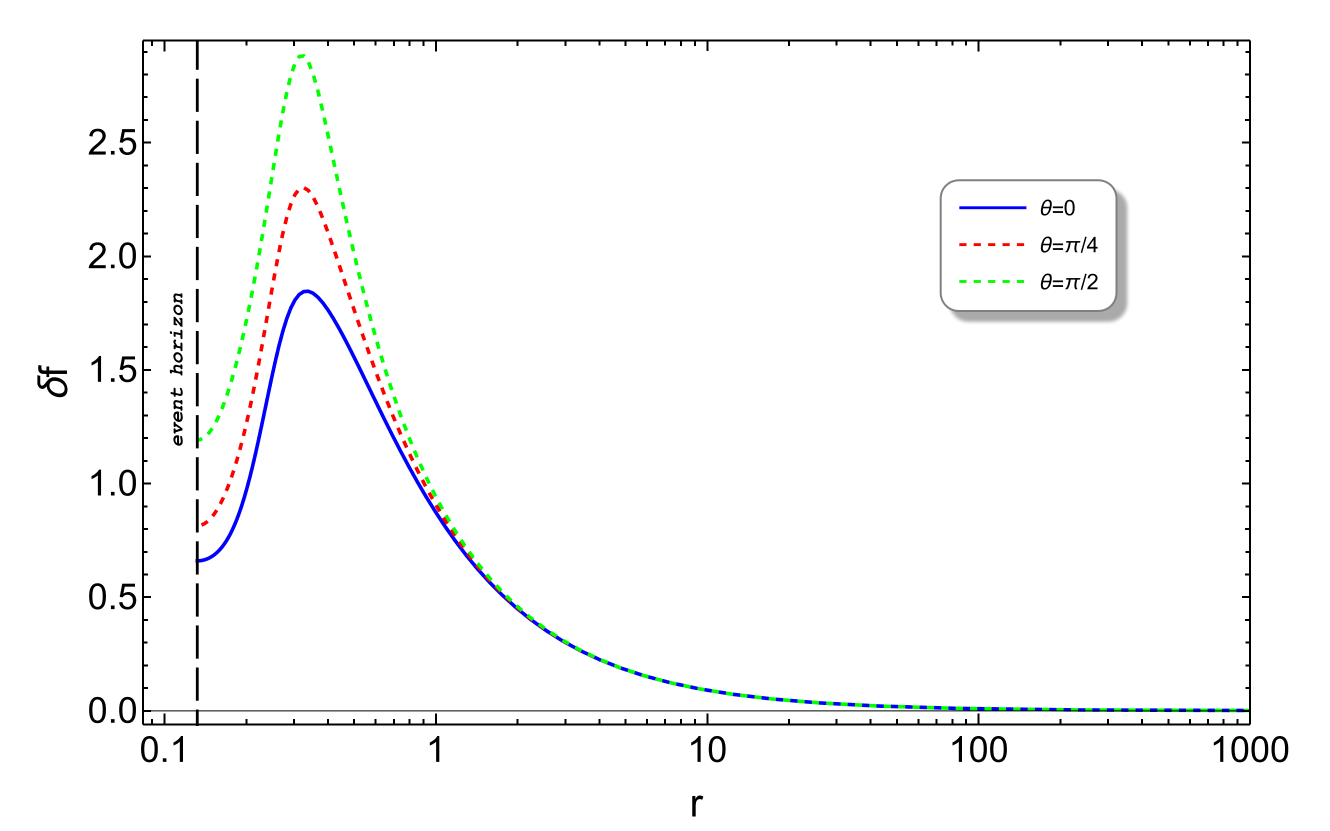}
    \end{minipage}
    \hspace{0.5cm}
    \begin{minipage}{0.4\textwidth}
	\centering
	\includegraphics[width=\textwidth]{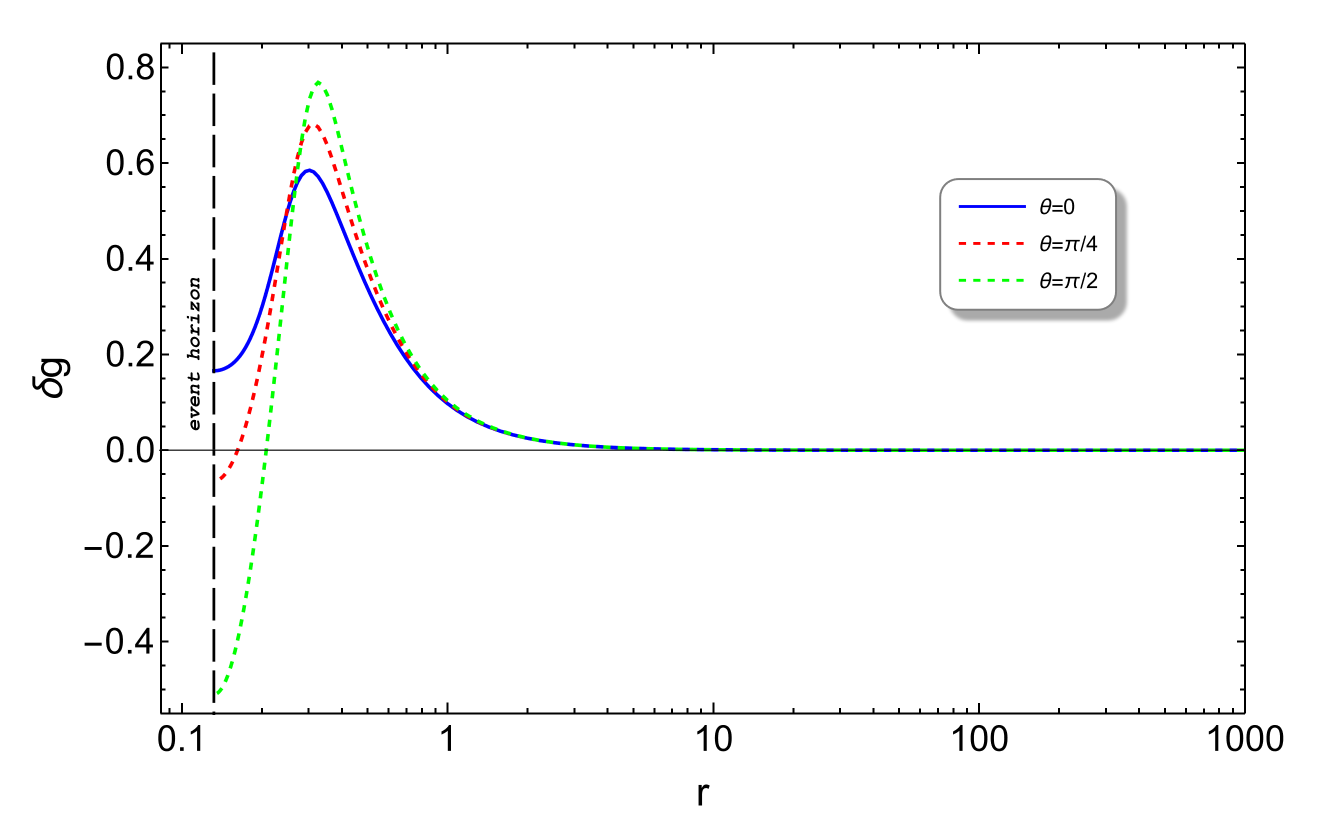}
    \end{minipage}
    
    \begin{minipage}{0.4\textwidth}
        \centering
	\includegraphics[width=\textwidth]{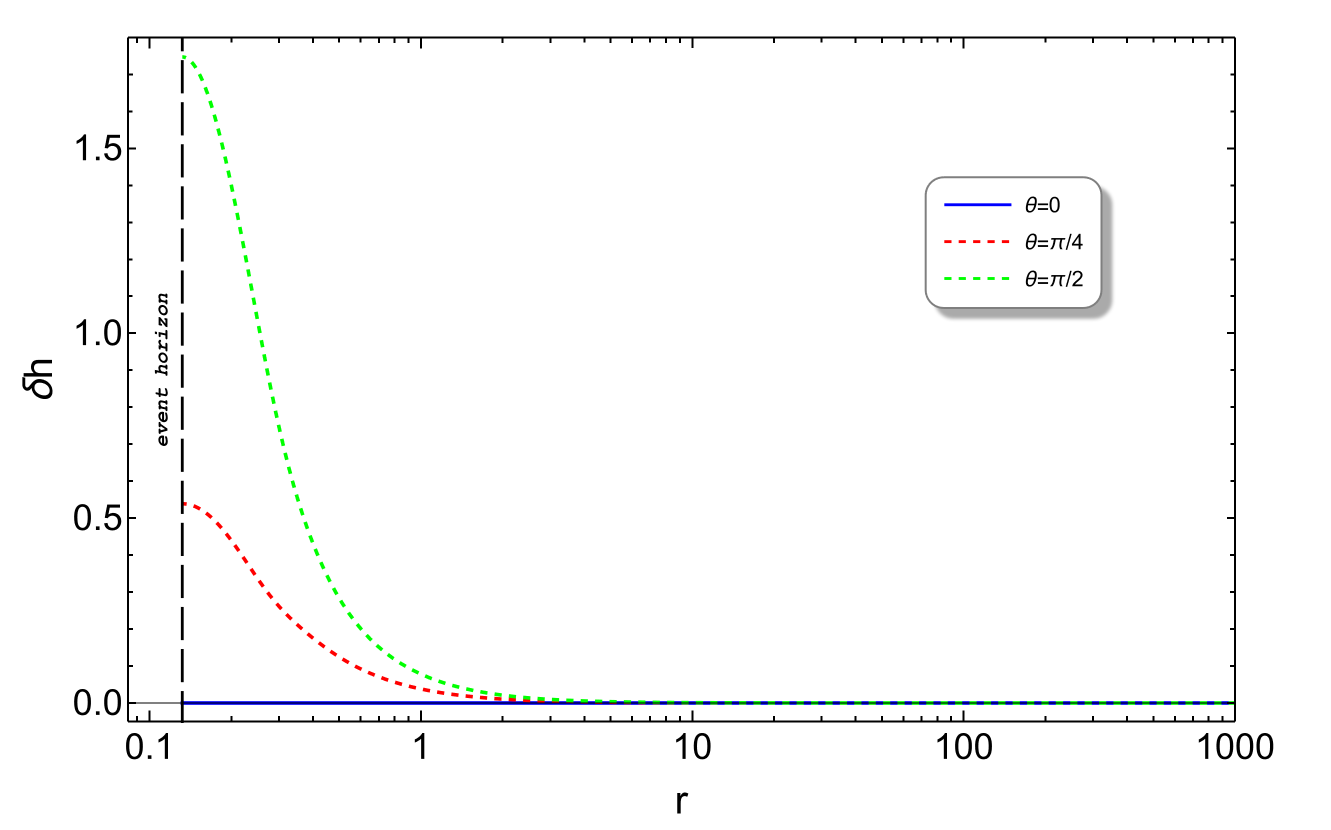}
    \end{minipage}
    \hspace{0.5cm}
    \begin{minipage}{0.4\textwidth}
	\centering
	\includegraphics[width=\textwidth]{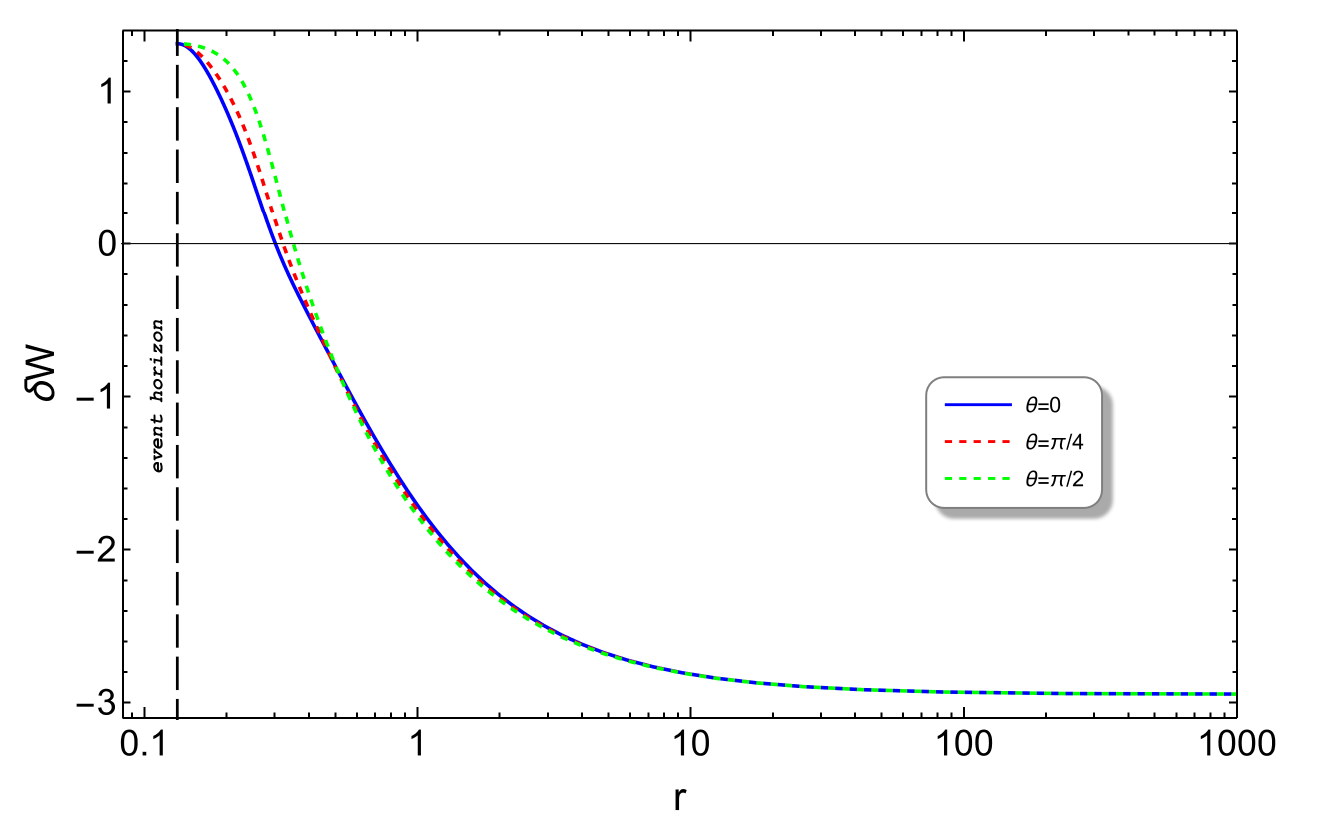}
    \end{minipage}

    \begin{minipage}{0.4\textwidth}
        \centering
	\includegraphics[width=\textwidth]{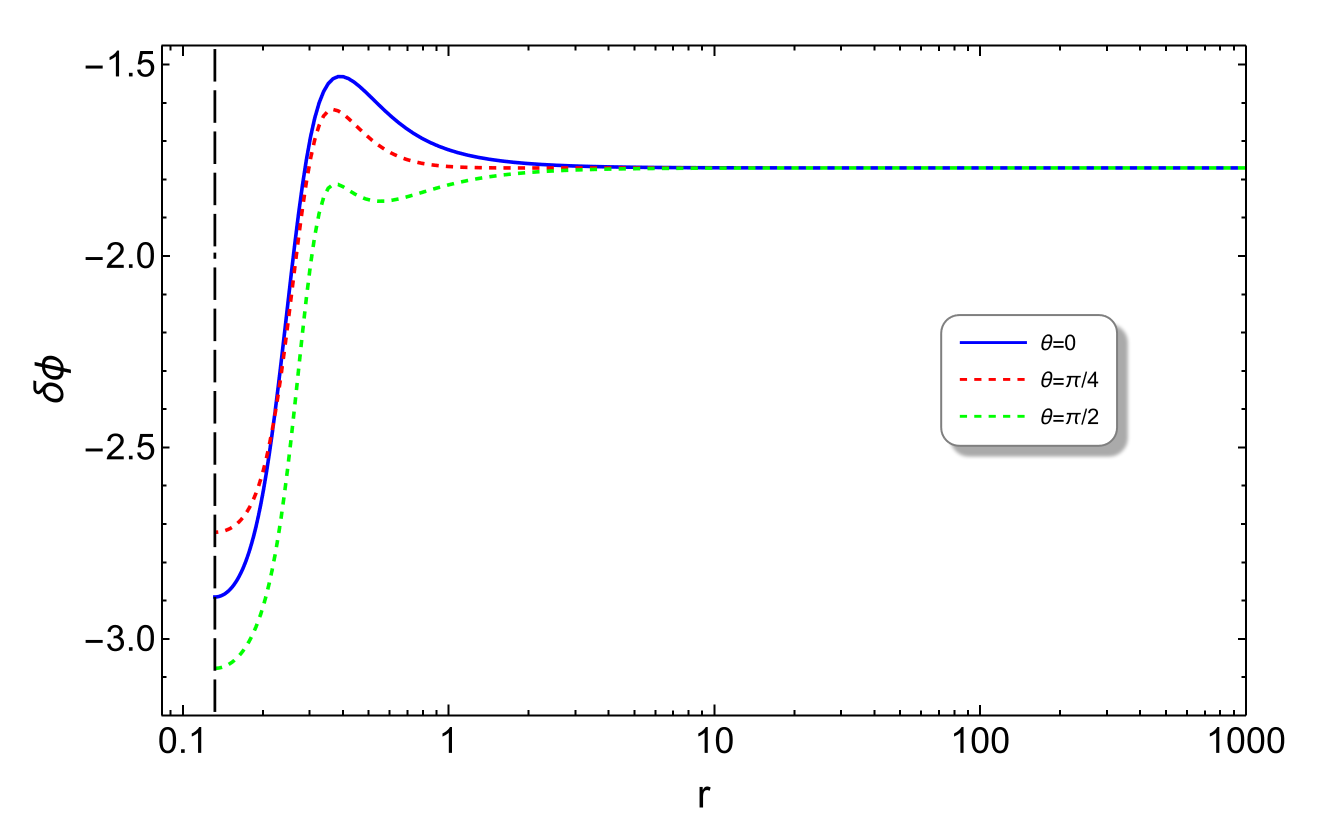}
    \end{minipage}
    \hspace{0.5cm}
  \begin{minipage}{0.4\textwidth}
	\centering
	\includegraphics[width=\textwidth]{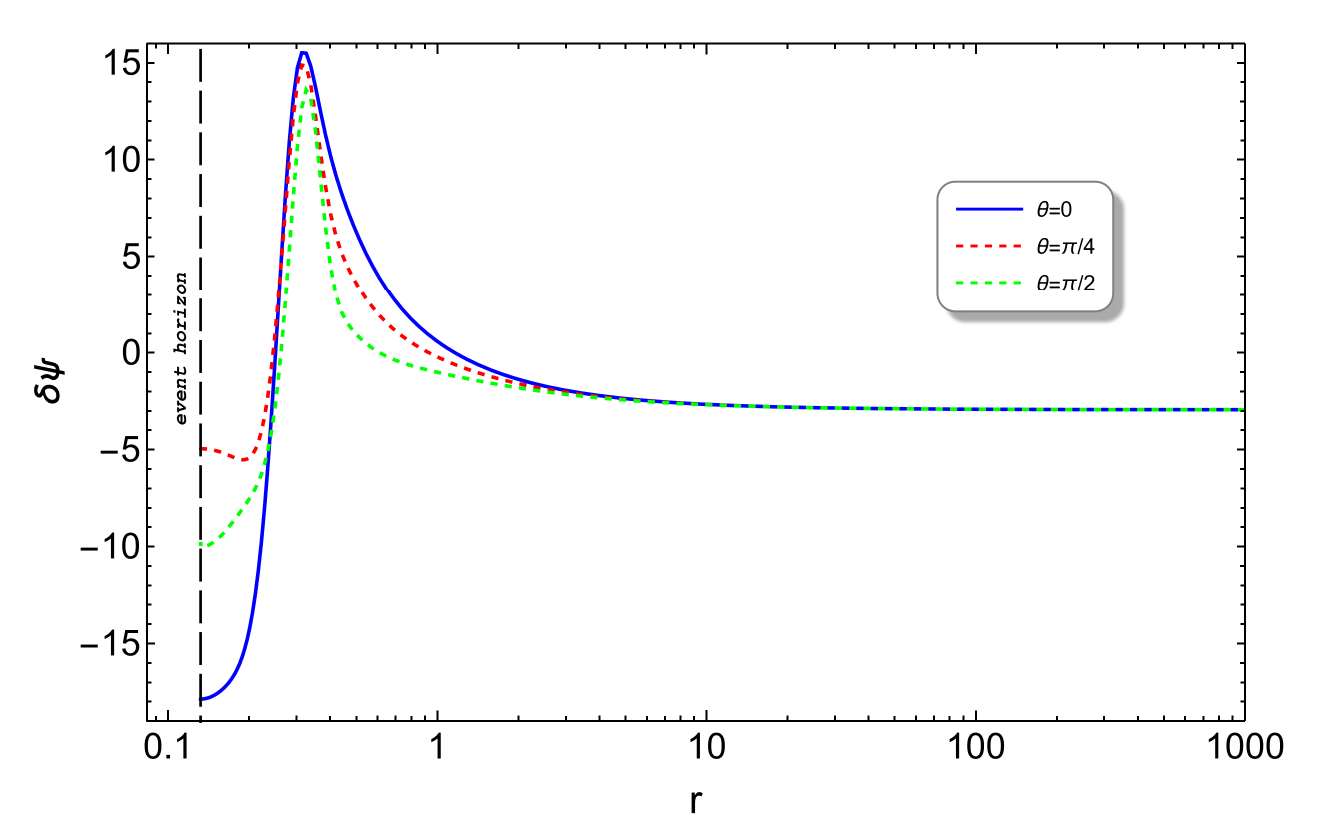}
    \end{minipage}

    \caption{Comparison of the metric functions $f,g,h,W$ and the scalar fields $\phi$ and $\psi$ for the nonlinearly scalarized rotating black hole solutions with coupling ratio $\alpha_2/\alpha_1=0$ and $\alpha_2/\alpha_1=0.1$, using the same parameters as Figs. \ref{fig:3D 2D metric functions}.}
    \label{fig:error}
\end{figure}

\begin{figure}[htbp]
    \centering
    \begin{minipage}{0.4\textwidth}
        \centering
	\includegraphics[width=\textwidth]{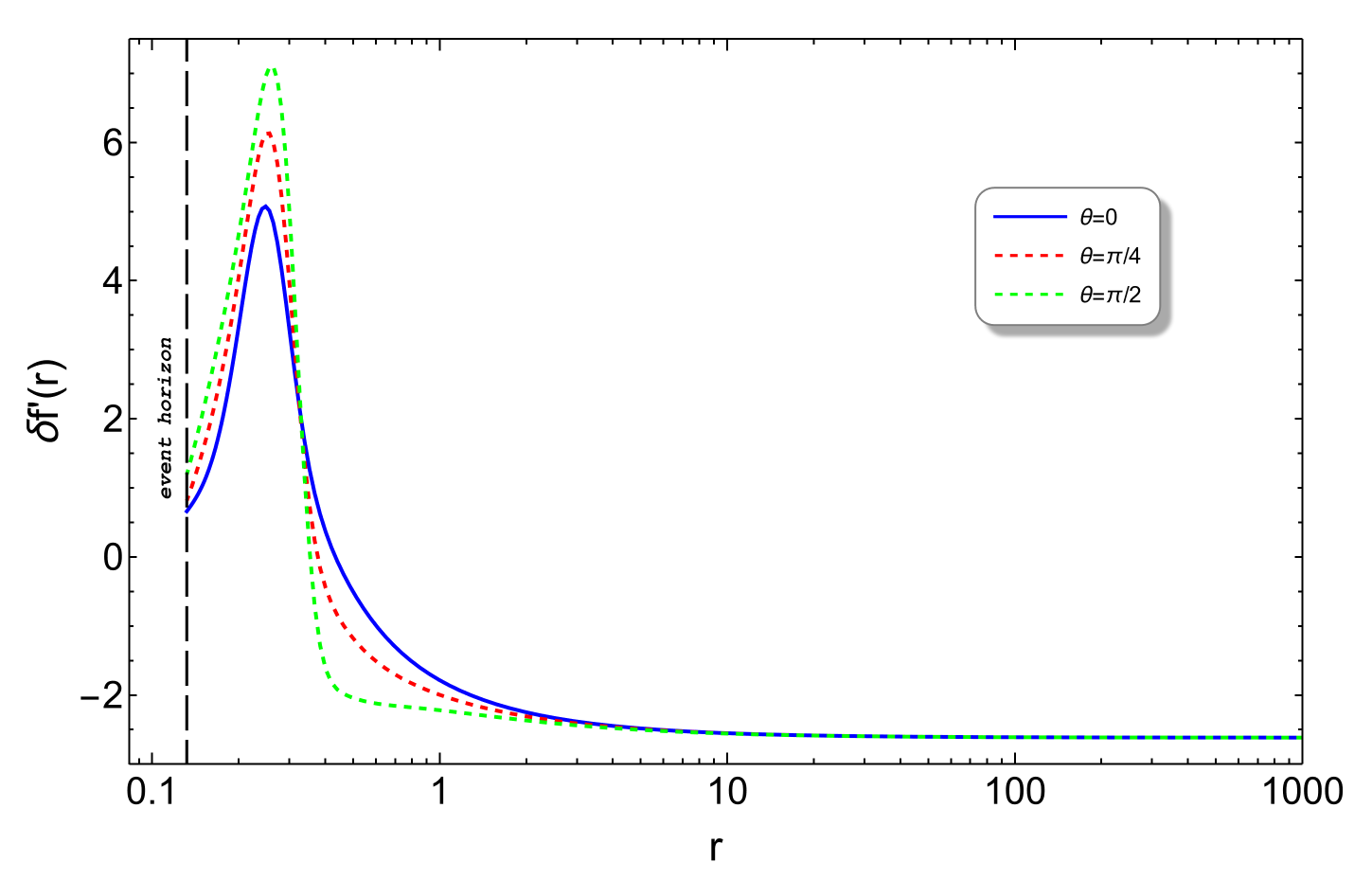}
    \end{minipage}
    \hspace{0.5cm}
    \begin{minipage}{0.4\textwidth}
	\centering
	\includegraphics[width=\textwidth]{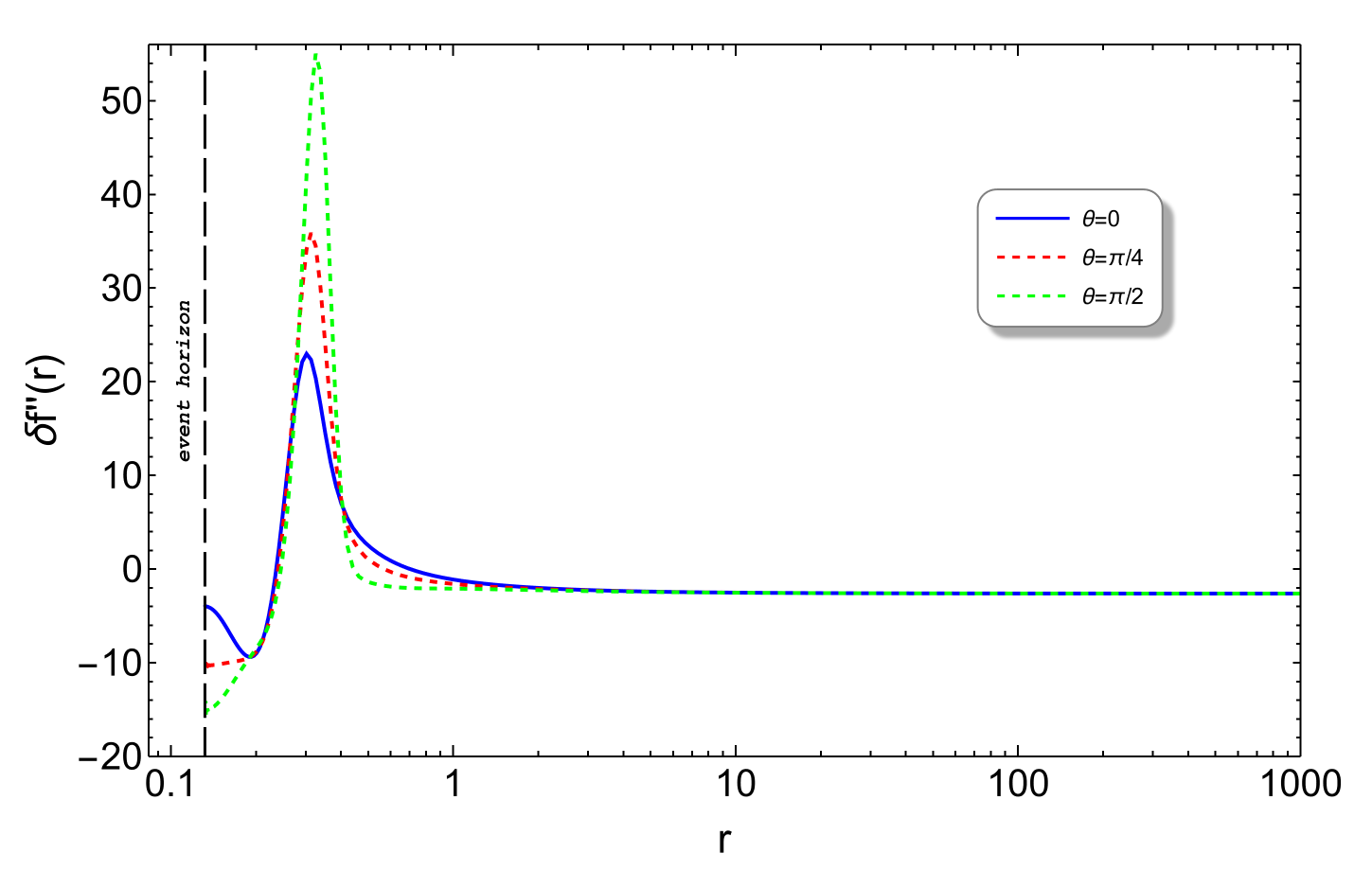}
    \end{minipage}
    \caption{The percentage change of the first order derivative (left panel) and second order derivative (right panel) of metric function $f$.}
    \label{fig:error df ddf}
\end{figure}

In order to explore how the quartic term $\mathcal{G}^2$ influences scalar fields and spacetime geometry, we compute the numerical solutions for the model with $\alpha_2/\alpha_1 = 0.1$ and compare them with those obtained with $\alpha_2/\alpha_1 = 0$. 
Set to the black hole parameters $\chi=0.4$,$r_H=0.132$, same as Figs. \ref{fig:3D 2D metric functions} and \ref{fig:3D 2D scalar fields} when $\alpha_2/\alpha_1 = 0.1$, allow us to calculate the scalar charge as $Q_s = 0.09435$ and the ADM mass as $M = 0.30485$. 
Using the formula $\delta \mathcal{F}^{(k)}=(\mathcal{F}^{(k)}_{\alpha_{2}=0.1}-\mathcal{F}^{(k)}_{\alpha_{2}=0})/\mathcal{F}^{(k)}_{\alpha_{2}=0}\times 100$, where $\mathcal{F}^{(k)} =\{f,g,h,W,\phi,\psi\}$, we quantify variations in each function's percentage change. 
In Fig. \ref{fig:error}, the outcomes of $\delta \mathcal{F}^{(k)}$  are displayed numerically. 
Except for $\psi$, the functions $\mathcal{F}^{(k)}$ only exhibit small deviations compared to when $\alpha_2/\alpha_1 = 0$, as observed. 
Around $0.1\%$ is the percentage change for the metric function $g$, while the deviation of other metric functions and the scalar field $\phi$ roughly under $3\%$ near the horizon. 
However, demonstrating a notable deviation, the scalar field $\psi$ (equivalent to $\mathcal{G}$) shows alterations up to $15\%$ near the horizon. 
Originating from minor scalar field variations, magnified by significant fluctuations in their radial second-order derivatives near the horizon, this substantial discrepancy arises. 
Depicted in Fig. \ref{fig:error df ddf}, as an example, is the percentage change in the radial derivatives of the metric function $f$. 

\section{Conclusions}
\label{Conclusions}
In this work, we have explored the non-linear instability of rotating black holes within a scalar Gauss-Bonnet gravity framework that includes an additional squared Gauss-Bonnet term. We adopted quartic-exponential coupling function $F(\phi)= \frac{1}{\kappa}(1-e^{-\kappa \phi^{4}})$, under which the Kerr black holes remain linearly
stable, however, sufficiently large perturbation can raise the nonlinear instability and lead to the formation of nonlinearly scalarized rotating black holes.

To elucidate the effect of the squared Gauss-Bonnet term $\mathcal{G}^2$ and the coupling parameter $\kappa$, we employed the Chebyshev pseudo-spectral method alongside the Newton-Raphson iterative approach to the coupled system of reduced field equations. Our analysis reveals that the effect of the $\mathcal{G}^2$ term is not universal but exhibits a striking dependence on $\kappa$, differing from the monotonically suppressive behavior reported in \cite{liu2025rotatingblackholesclass}.
For the strong coupling case ($\kappa = 6$), the $\mathcal{G}^2$ term induces a non-monotonic modification of the static black hole existence domain, with the most extended mass range occurring at $\alpha_2=0.1$. In contrast, for intermediate and weaker coupling ($\kappa=25$ and $\kappa=100$), a monotonically suppressive effect is observed, where the domain consistently contracts with increasing $\alpha_2$. Furthermore, a numerical investigation of the $\kappa=6$ case was conducted to explore the possibility of highly rotating scalarized solutions. Our attempts yield convergent solutions with spins up to $\chi \approx 0.4$ when $\alpha_2 \neq 0$. For the $\alpha_2 = 0$ case, however, our searches did not converge to stable solutions beyond $\chi \approx 0.1$ under the same numerical setup. This outcome may indicate a potential facilitative role for the $\mathcal{G}^2$ term in high-spin regime, which warrants further investigation.

Regarding the thermodynamic properties, our analysis of the Helmholtz free energy reveals a first-order phase transition between rotating black holes and the Kerr black holes. For a fixed spin parameter $\chi$, there exists a critical Hawking temperature $T_c$: when $T_H < T_c$, the Kerr black hole represents the thermodynamically stable ground state, whereas for $T_H > T_c$, the scalarized black hole exhibits lower free energy and becomes thermodynamically preferred. This critical temperature increases with the spin parameter $\chi$, indicating that the scalarized phase transition can occur over a broader temperature range for rapidly rotating black holes. In particular, for intermediate coupling cases (e.g. $\kappa=25$), the entropy of branch 1 scalarized black holes is higher than that of Kerr black holes across most the mass range, further confirming their thermodynamic preference. In contrast, both the entropy and free energy of branch 2 solutions consistently indicate thermodynamic instability relative to the Kerr metric. This thermodynamic behavior exhibits a strong dependence on the coupling strength $\kappa$, highlighting the rich phase structure manifested by the nonlinear scalarization mechanism in strong-field gravity.

Future research offers several opportunities to extend these findings. Our findings indicate that the squared Gauss-Bonnet term, and the coupling parameter $\kappa$ significantly affect the scalarization domain, suggesting that investigating alternative coupling functions or exploring extreme
values of $\kappa$ (e.g. approaching zero or infinity) could reveal new scalarized solutions or elucidate the broader applicability of these effects. Additionally,
to connect these theoretical solutions to astrophysical observations, calculating observable properties, such as gravitational wave quasi-normal modes or black hole 
shadow profiles, could distinguish scalarized black holes from Kerr black holes, potentially enabling their detection with instruments like Event Horizon Telescope or future 
gravitational wave observatories. Exploring these directions would deepen our understanding of scalarized rotating black holes in this modified gravity framework and
enhance their significance for astrophysical and theoretical studies.

\section*{ACKNOWLEDGMENTS}
This research is supported by the National Natural Science Foundation of China under Grant Nos.$12375056$, $12375048$, 
and the Postgraduate Research $\&$ Practice Innovation Program of Jiangsu Province under Grant No.KYCX$24\_3712$.
Some of our calculations were performed using the tensor-algebra bundle xAct~\cite{xact}.

\begin{appendices}
{\centering \section{Appendix A: Resolution Settings for Computation}}
\label{Appendix A}

In this appendix, we explain why we primarily set the resolution as $N_{x} = 40$ and $N_{\theta}=8$ when using the Chebyshev pseudo-spectral method to solve the PDEs.
\begin{figure}[htbp]
    \centering
    \begin{minipage}{0.45\textwidth}
        \centering
	\includegraphics[width=\textwidth]{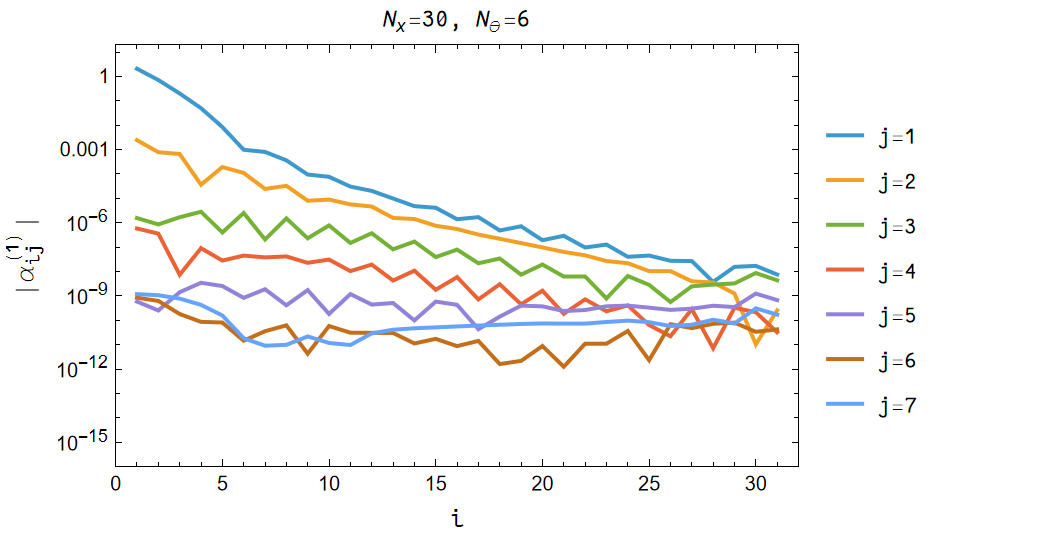}
    \end{minipage}
    \hspace{0.1cm}
    \begin{minipage}{0.45\textwidth}
	\centering
	\includegraphics[width=\textwidth]{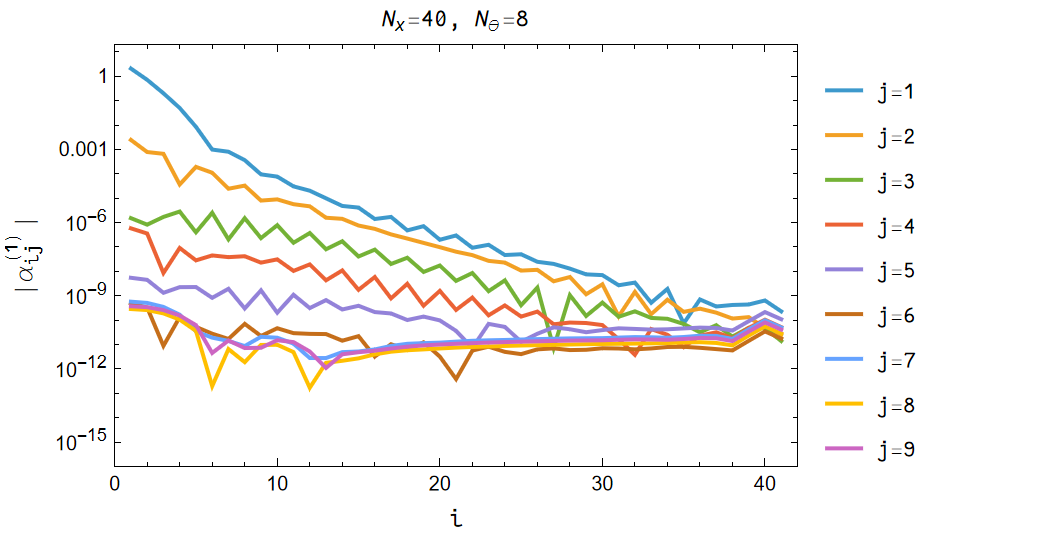}
    \end{minipage}
    
    \begin{minipage}{0.45\textwidth}
        \centering
	\includegraphics[width=\textwidth]{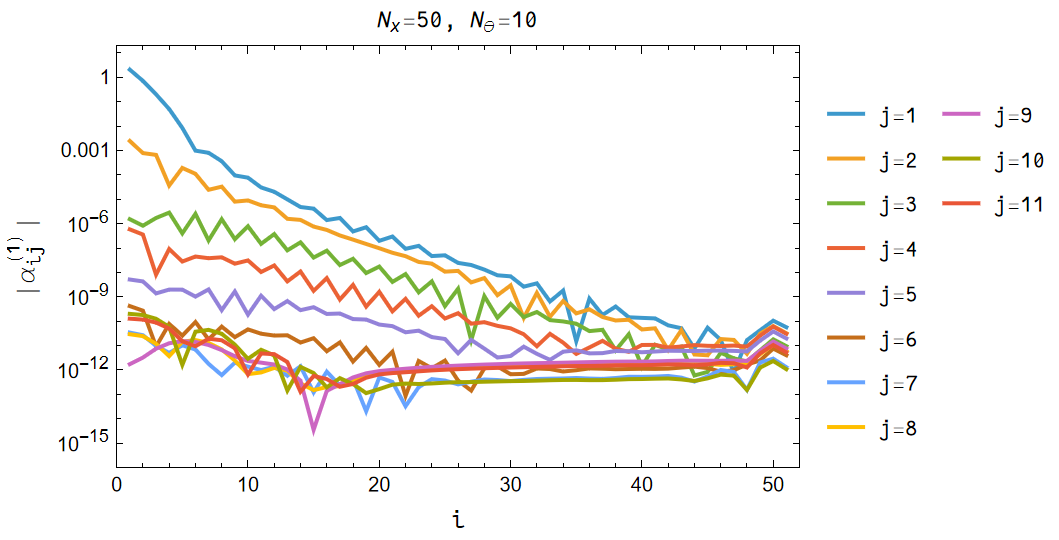}
    \end{minipage}
    \hspace{0.1cm}
    \begin{minipage}{0.45\textwidth}
	\centering
	\includegraphics[width=\textwidth]{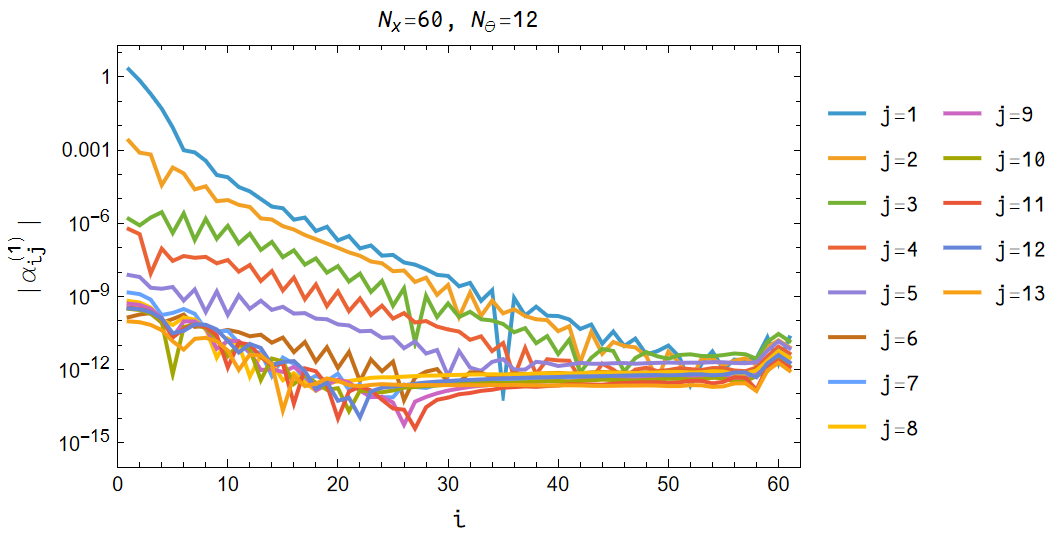}
    \end{minipage}
    
    \caption{
    The absolute values of the spectral decomposition coefficients $\alpha^{(1)}_{ij}$ of the metric function $f\left(x, \theta\right)$ with different $\left(N_x, N_{\theta}\right)$.
    }
    \label{fig:coefficient metric}
\end{figure}

Recall that in Sec. \ref{Numerical method and quantities of interest}, the ultimate purpose of the numerical approach is to find the spectral coefficients $\alpha_{ij}^{k}$, as expressed in Eq. (\ref{Coeffs}). For example, we randomly select a set of input parameters for scalarized black hole, $\kappa = 100,~\alpha_2 = 0.1,~r_H = 0.144$ and $\chi = 0.2$, and solve the PDEs using the various resolutions. Here, we focus on exploring the magnitude of the coefficients of the first metric function $f(x,\theta)$, specifically $\big|\alpha_{ij}^{(1)}\big|$. As illustrated in Figs. \ref{fig:coefficient metric} for this nonlinearly scalarized rotating black hole solution, the magnitudes of $|\alpha_{ij}^{(1)}|$ exhibit an exponential decay with increasing index $i$, demonstrating the convergence of the numerical scheme. Consequently, the functional value is predominantly determined by the leading terms in the $\alpha_{ij}^{k}$ series. Furthermore, as shown in Figs. \ref{fig:coefficient metric}, when we set $N_{x} = 30$ and $N_{\theta}=6$,it is uncertain whether these leading terms sufficiently approximate the true scalarized black hole solution. As the number of nodes increases, the trailing terms stabilize to a small value of approximately $10^{-12}$. Considering both the computational efficiency and numerical accuracy, we choose $N_{x} = 40,~N_{\theta}=8$ as our main resolution for computations in this work. Additionally, we enhance the resolution at the boundary of the existence domain to ensure accuracy, such as $\chi=0$ the non-rotating scalarized black hole.

{\centering \section{Appendix B: Supplementary Data for $\kappa = 6$}}
\label{Appendix B}

\begin{table}[htb]
    \centering
    \caption{The mass range $\Delta M / \sqrt{\alpha_1}$ for scalarized black holes with $\kappa=6$ and varying coupling constant $\alpha_2$ and dimensionless spin $\chi$.}
    \label{tab: kappa=0.6_alpha_2=all}
    \begin{tabular}{lcccccc}  
        \toprule 
        & \multicolumn{5}{c}{Dimensionless Spin $\chi$} \\ 
        \cmidrule(lr){2-6} 
        Coupling $\alpha_2$       & 0 & 0.1 & 0.2 & 0.3 & 0.4 \\ 
        \midrule 
        \specialrule{0.5pt}{0em}{0em}  
         0.1            & 0.69187 & 0.67805 & 0.65897 & 0.63052 & 0.57585   \\
        \specialrule{0.5pt}{0em}{0em}  
         0.2            & 0.55970 & 0.45991 & 0.44413 & 0.40000 & 0.37010  \\
        \specialrule{0.5pt}{0em}{0em}  
         0.3            & 0.58655 & 0.40896 & 0.38398 & 0.32123 & 0.14886  \\
        \bottomrule 
    \end{tabular}
\end{table}

\end{appendices}

\centering
\bibliographystyle{unsrt}
\bibliography{BmyRef.bib}

\end{document}